\documentclass[aps,pre,epsf,superscriptaddress,amsmath,amssymb,amsfonts,  twocolumn, showpacs]{revtex4-1}

\usepackage{color} 
\newcommand{\abs}[1]{\left| #1 \right|}

\usepackage[colorlinks=true,
linkcolor=blue,
filecolor=magenta,      
urlcolor=blue,
citecolor=blue]{hyperref}
\usepackage{graphicx,float}
\usepackage{amsmath}
\usepackage{physics}
\usepackage{amsfonts}
\usepackage{amssymb}
\usepackage{times}
\usepackage{bm}
\everymath{\displaystyle}
\begin{document}
\title{ Quench induced vortex-bright-soliton formation\\ in binary Bose-Einstein condensates}

\author{K. Mukherjee}
\affiliation{Indian Institute of Technology Kharagpur, Kharagpur-721302, West Bengal, India}
\affiliation{Center for Optical Quantum Technologies, Department of Physics, University of Hamburg, 
Luruper Chaussee 149, 22761 Hamburg Germany} 
\author{S. I. Mistakidis}
\affiliation{Center for Optical Quantum Technologies, Department of Physics, University of Hamburg, 
Luruper Chaussee 149, 22761 Hamburg Germany}
\author{P. G. Kevrekidis}
\affiliation{Department of Mathematics and Statistics, University of Massachusetts Amherst, Amherst, MA 01003-4515, USA}
\affiliation{Center for Optical Quantum Technologies, Department of Physics, University of Hamburg, 
Luruper Chaussee 149, 22761 Hamburg Germany}  
\author{P. Schmelcher}
\affiliation{Center for Optical Quantum Technologies, Department of Physics, University of Hamburg, 
Luruper Chaussee 149, 22761 Hamburg Germany}\affiliation{The Hamburg Centre for Ultrafast Imaging,
University of Hamburg, Luruper Chaussee 149, 22761 Hamburg,
Germany}

\date{\today}


\begin{abstract}

We unravel the spontaneous generation of vortex-bright-soliton structures in binary Bose-Einstein condensates with a small mass imbalance 
between the species confined in a two-dimensional harmonic 
trap where one of the two species has been segmented into two parts by a potential barrier. 
To trigger the dynamics the potential barrier is suddenly removed and subsequently the segments perform a counterflow dynamics. 
We consider a relative phase difference of $\pi$ between the segments, while a singly quantized vortex may be imprinted at the center of the other species. 
The number of vortex structures developed within the segmented species following the merging of its segments is found to depend on the presence of an 
initial vortex on the other species. 
In particular, a $\pi$ phase difference in the segmented species and a vortex in the other species result in a single vortex-bright-soliton structure. 
However, when the non-segmented species does not contain a vortex the counterflow dynamics of the segmented species gives 
rise to a vortex dipole in it accompanied by two bright solitary waves
arising in the non-segmented species.
Turning to strongly mass imbalanced mixtures, with a heavier segmented species, we find that the same overall dynamics takes place, 
while the quench-induced nonlinear excitations become more robust. 
Inspecting the dynamics of the angular momentum we show that it can be transferred from one species to the other, and its transfer rate can be tuned 
by the strength of the interspecies interactions and the mass of the atomic species.  
\end{abstract}

\pacs{67.85.-d, 67.40.Vs, 67.57.Fg, 67.57.De}
\maketitle


\section{Introduction}\label{Introduction}

The experimental realization of Bose Einstein condensates (BECs) \cite{Anderson198, Davis1995} has opened up a very promising venue to explore a 
variety of non-linear excitations, such as quantized vortices \cite{Marzlin1997, Dalfovo1996, Rokhsar1997, Dodd1997, Jackson1998}, that 
arise in such an environment (see also
the reviews~\cite{Alexander2001,our2}). 
Vortices have been a long-standing theme of interest in a wide range of branches of physics, ranging from optics \cite{EWolf2009} and classical hydrodynamics \cite{Lugt1983} 
to cosmology \cite{Noam2013} and condensed-matter physics \cite{Blatter1994}, and are intimately connected to the phenomenon of Bose-Einstein condensation~\cite{Pismen1999}. 
Most importantly, they constitute an ideal testbed for studying superfluidity \cite{Onofrio2000} and quantum turbulence \cite{Paoletti2008, White2010, Neely2013, Kwon2014}. 
State-of-the-art BEC experiments enable us to routinely create \cite{Matthews2498,Anderson2000} such topological structures and inspect their equilibria \cite{Feder2001,Cooper2004,Isoshima1999} 
and dynamical properties \cite{chevy2003, Coddington2003, Baym2003, Mizushima2004} with an unprecedented level of accuracy \cite{Freilich2010,Anderson2010,Halltorres}. 
Moreover, a multitude of studies have focused on the creation of multiple vortices of the same~\cite{Madison2000,Halltorres} or
of opposite charges~\cite{Middelkamp2011,bettina1},
vortices of higher topological
charge \cite{Shibayama2011,Jukka2018, Leanhardt2002,Kumakura2006} 
and the associated dynamical instabilities of these topological objects \cite{Shin2004_dyna, Mateo2006}. 

Different widely used techniques to generate vortices, include for example, phase imprinting \cite{Andersen2006,Wright2008, Wright2009}, rotation of 
the external trap \cite{Hodby2001, Williams2010,Hodby2003} or the presence of synthetic magnetic fields \cite{Murray2009,Zhao2015, Price2016}. 
Moreover, processes involving dynamical instabilities~\cite{Anderson2001}
or the quenching of a system parameter \cite{Anglin1999, Weiler2008}  can also result in topological defect 
formation. 
In this latter context, macroscopic interference of multiple slices of a BEC confined in an external trap \cite{Andrews1997,Myatt1997,Pitaevskii1999,Liu2000,Hadzibabic2004}, bears close 
resemblance with the celebrated Kibble-Zurek mechanism \cite{Kibble1976, Zurek2009}. 
In particular, in an elongated three-dimensional trap, the interference of BEC clouds can give rise to the formation of dark solitary waves, and subsequently the nucleation of arrays 
of vortex rings. 
The spontaneous formation of vortices has also been experimentally observed \cite{Scherer2007} during the
interference of three individual BEC components. 
The impact of the finite removal time of the potential barrier and the phase difference between the different BEC slices on the formation of vortices has also been 
investigated \cite{Wallis1997, Anderson2008}. 

Although such  interference processes have been examined extensively for single species \cite{Scott1998, Wallis1997, Xiong2013, Yang2013}, the corresponding scenario is far less explored for 
two dimensional (2D) multispecies BECs. 
In the latter context, a variety of non-trivial patterns can arise.
These include square vortex lattices \cite{Mueller2002, cornell_vl, Pekko2012}, triangular vortex lattices \cite{Mueller2002}, serpentine vortex sheets \cite{Kasamatsu2009}, 
striated magnetic domains \cite{Miesner1999}, robust target patterns \cite{Mertes2007}, and longitudinal spin waves \cite{Lewandowski2002}. 
Also, various nonlinear matter wave structures, such as dark-bright
solitons~\cite{Becker2008},
and bound states thereof~\cite{engels1,engels3}, dark-dark solitons \cite{Santos2001,engels2}, soliton 
molecules \cite{Vadym2003}, and symbiotic solitons \cite{Belmonte2005} can occur in multispecies BECs (see also~\cite{KEVFRA} for a relevant review).
Many of these stem from effectively one-dimensional settings.
On the other hand, the decay of dark soliton stripes to chains of vortices with opposite topological charges has been vastly addressed in atomic 
condensates \cite{Anderson2001,Kevrekedis2004, Ma2010}. 
Here, stability analysis of dark soliton stripes revealed that such states are prone to decay via the so-called snake instability, i.e. instability against transverse 
long-wavelength perturbations \cite{Kuznetsov1995}. 
Also very recently, the existence, stability and dynamics of more involved two species dark-bright soliton stripes leading to the formation of vortices
coupled with bright solitons in the second component (the so-called
vortex brights (VBs)~\cite{laww})
via the aforementioned 
modulation instability have been examined using an adiabatic 
invariant approach \cite{kevrekidis2018}. 

Motivated by the above-mentioned studies we hereby attempt to interweave two
settings, namely, interference of BECs as a means of inducing
spontaneous pattern formation
and the 2D multicomponent BECs as a rich platform where different types
of resulting patterns are available. 
Moreover, the existence of the vortex in one of the species can give rise to angular momentum exchange processes 
between the species due to the presence of the interspecies coupling. 
Certainly, the control of angular momentum transfer would be also highly desirable.
Our starting point are two weakly interacting BECs, alias species A and species B, which are harmonically confined in a quasi 2D geometry. 
The external trap of species B is initially partitioned into two parts by a repulsive potential barrier at the trap center, thus splitting species B into two segments. 
After preparing the initial state of the system the potential barrier of species B is 
ramped down inducing a counterflow dynamics for this species. 
We focus on a slightly mass imbalanced mixture. 
Note that the induced counterflow dynamics will be qualitatively similar for mass balanced binary BECs, being experimentally realizable using two hyperfine states 
of the same atomic species. However, in this case, the experimental realization of a potential barrier experienced by only one hyperfine state might be 
experimentally challenging.

We consider two different initial configurations of species A. 
In the first configuration, species A contains a singly quantized vortex, and in the second case, this species is in its ground state configuration 
(i.e. a state with zero vorticity). 
On the other hand, a $\pi$ relative phase difference is imprinted in the segments of species B. 
In both cases the merging of the two segments in species B creates soliton stripes which, due to their transverse instability, will break into 
vortex-antivortex (VAV) pairs \cite{PhysRevA.62.053606}. 
In turn these entities evolve into non-trivial topological structures such as a VAV pair known as dipole \cite{Kevrekedis2004}, 
a VAV pair and a vortex (VAV-V) \cite{Seman2010}, or two dipoles
\cite{yang2016dynamics}. This pattern formation in the two components
and
its triggering by the quench-induced interference is one of the
principal
scopes of the present work. 
Depending on the initial configuration, distinct features in the long-time dynamics of the system arise. 
For the scenarios including trap parameters and chemical
potentials considered herein, in the first configuration, a single vortex is created in species B coupled to a bright soliton in species A. 
This composite structure will be referred to as  a VB  soliton~\cite{laww}
hereafter.
In fact, the above-mentioned composite structures have been referred to with different names
by different communities including VB solitons~\cite{laww},
baby Skyrmions~\cite{sutcliffe} and filled-core vortices~\cite{skryabin}. 
We will stick to the former notation here. 
Remarkably enough, a
transfer of angular momentum between the species takes place in the course of the dynamics, 
with a transfer rate strongly dependent on the interspecies interaction strength. 
Turning to the second configuration, a vortex dipole is created in species B and two bright solitons in species A, while 
no net angular momentum is transferred from species A to species B. 
Finally, exploiting a stronger mass imbalanced mixture, where species B possesses the larger mass, gives rise to a similar to 
the above-described overall dynamics, but the quench-induced non-linear structures are more robust in time and a more efficient angular momentum transfer occurs. 

The remainder of this work is structured as follows. 
In Sec.~\ref{The Model} we describe our setup and discuss the observables which we utilize to monitor the dynamics of the system. 
We discuss the nonequilibrium dynamics of the binary bosonic system consisting of slightly mass imbalanced components focusing on 
the configuration with a singly quantized vortex in Sec.~\ref{Equal_Mass} and no vortex in species A in Sec.~\ref{Discussion2}. 
The effect of strong mass imbalance between the involved components on the dynamics is described in Sec.~\ref{unequal_Mass}.
We summarize our results and provide possible future extensions in Sec.~\ref{conclusion}. 
Appendix \ref{Imperfection} presents the quench induced dynamics in the case that the two segments of species B initially have a 
particle imbalance or they are not symmetrically placed with respect to the trap center.  
Finally, in Appendix \ref{Derivation} we briefly outline the derivation of the equations of evolution of the angular momentum of each species.


\section{Theoretical background}\label{The Model}

\subsection{The Setup}\label{hamiltonian}

We consider a binary BEC of two different atomic species prepared in a 2D external trapping potential 
The dynamics of the system is governed, in the mean-field approximation, by the following set of coupled Gross-Pitaevskii (GP) equations
\begin{equation}\label{1}
\begin{split}
&i\hbar \frac{\partial  \Psi_{\sigma} (\textbf{r, t}) }{\partial t} \\& = \left[-\frac{\hbar^2\nabla^2}{2m_{\sigma}}+ V_{\sigma}(\textbf{r})+  \sum_{\sigma'}^{} U_{\sigma \sigma'}|\Psi_{\sigma'}
(\textbf{r}, t)|^2 \right ] 
\Psi_{\sigma}(\textbf{r}, t)  .
\end{split}
\end{equation}
The index $\sigma = A, B$ refers to each of the species of the binary BEC, while $m_{\sigma}$ and $\psi_{\sigma}(\textbf{r}, t)$ denote the mass and the condensate wavefunction 
of the $\sigma$ species respectively. 
Within the ultracold regime, $s$-wave scattering is the dominant interaction process and therefore the interparticle interactions can be modeled by contact interactions. 
The intraspecies coupling constants $ U_{\sigma\sigma}=2\pi \hbar a_{\sigma \sigma}/m_{\sigma}$ are characterized by the scattering lengths $a_{AA}$ and $a_{BB}$. 
The corresponding interspecies scattering coupling is $U_{AB} = 2 \pi \hbar a_{AB}/m_{AB}$, with $m_{AB} = m_A m_B/(m_A+m_B)$ being the reduced mass. 
$U_{AB}$ is determined by the scattering length $a_{AB}$ where an atom of one species scatters from an atom of the other species.
We assume a harmonic oscillator potential, i.e. $V_{\sigma} (\textbf{r}) = V_{\sigma}(x,y,z) = m_{\sigma} \omega^2_{\sigma} (x^2 +\alpha_{\sigma}^2 y^2 + \lambda_{\sigma}^2 z^2 )/2$ 
as the external trapping potential. 
Here, $\omega _{\sigma} $ denotes the trapping frequency along the $x$ direction, while $\alpha_{\sigma}$ and $\lambda_{\sigma}$ are the two anisotropy parameters. 
In this study, we mainly consider the case where the atoms of both components experience the same trapping frequency. 
Hence $\omega_A = \omega_B = \omega_x$, $\alpha_A = \alpha_B = \alpha = \omega_y/\omega_x$, and $\lambda_A = \lambda_B = \lambda = \omega_z/\omega_x$.
For the length and the energy scales of the system we choose the harmonic oscillator length $a_{\rm osc} = \sqrt{\hbar/m_A \omega_x}$ and the energy quanta $\hbar \omega_x$ respectively. 
Then we cast the above coupled set of GP equations into a dimensionless form by scaling the spatial coordinates as 
$x' = x/a_{\rm osc}$, $y' =y/a_{\rm osc}$ and $z' = z/a_{\rm osc}$, 
the time as $t'= \omega_x t$ and the wavefunction as $\Psi'_{\sigma}(x', y', z') = \sqrt{a_{\rm osc}^3/N_{\sigma}} \Psi_{\sigma}(x, y, z, t)$. 
For simplicity, in the following we drop the primes. 

\begin{figure}[ht]
\centering
\includegraphics[width=0.45\textwidth]{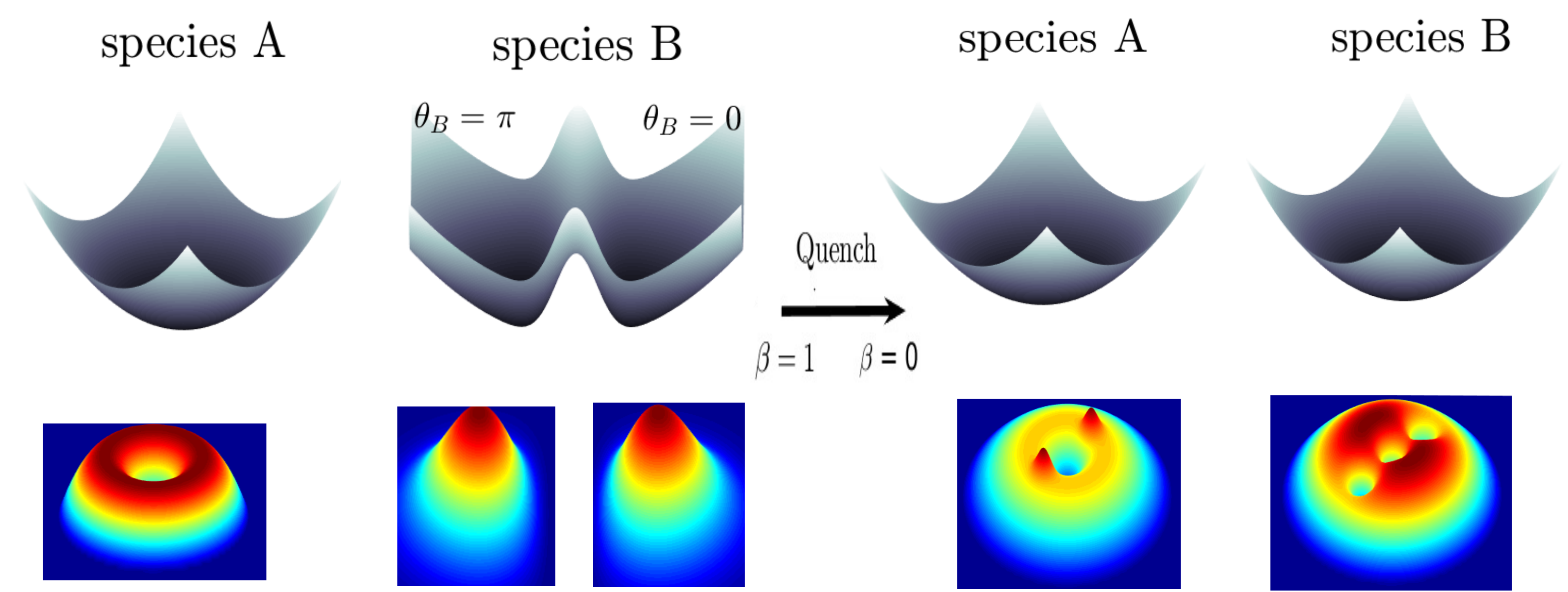}
\caption{Schematic representation of the considered setup and the quench protocol. 
A vortex of unit charge is initially imprinted at the trap center of species A and a $\pi$ relative phase difference 
between the segments of species B. 
At $t=0$ we suddenly ramp down the potential barrier of species B, thus quenching its trapping geometry from a 2D spatially anisotropic harmonic trap to 
an isotropic one (see also Sec. \ref{hamiltonian}). 
This process leads to the generation of vortices in species B and density humps in species A.} 
\label{fig:0}
\end{figure} 

For our purposes, after preparing the two BECs in a 2D harmonic potential, we use a repulsive barrier in the spirit of~\cite{Scherer2007,Whitaker2008}
to segregate species B into 
two isolated segments [see Fig. \ref{fig:0}]. 
The potential barrier reads $V_L = \beta(t)U_0e^{-\frac{x^2}{2a^2}}$, with $U_0$ and $a$ being the height and the width parameters of the potential barrier respectively.  
Initially, $\beta(t) = 1$ and depending on the functional form of $\beta(t)$ in $V_L$, various quench protocols of the barrier height can be realized. 
For instance, if $\beta(t)=1$ at $t<0$ and $\beta (t) = 0$ for $t \geq 0$,
this leads to a sudden quench [see Fig. \ref{fig:0}], while e.g. $\beta (t) = (t_f - t)/t_f$ 
refers to a time-dependent removal of the barrier within a given ramp-down time $t_f$ which is not considered herein. 
Moreover, due to the the quasi-2D geometry i.e. $\omega_z  \gg \omega_x, \omega_y$ of the potential, the wavefunction of each species can be factorized as follows  
\begin{equation}\label{2}
\Psi_{\sigma} (x, y, z, t) = \psi_{\sigma} (x, y, t) \phi_{\sigma} (z).
\end{equation} 
Here, $\phi_{\sigma} (z) $ is the normalized ground state wavefunction in the $z$ direction. 
Then the dimensionless form of the coupled GP equations of Eq. (\ref{1}) after integrating 
over $\phi_{\sigma}(z)$, since we address a 2D system, is the following \cite{Bandyopadhyay2017} 
\begin{equation}\label{3}
\begin{split}
i\frac{\partial \psi_A(x,y,t)}{\partial t} = &\bigg[- \frac{1}{2 } \nabla^{2}_{\perp}+ V_A(x,y) \\ &+ \sum_{j=A,B}^{} \mathcal{U}_{Aj} |\psi_j(x, y, t)|^2 \bigg ] \psi_A(x,y,t)
\end{split} 
\end{equation}
and 
\begin{equation}\label{4}
 \begin{split} 
i\frac{\partial \psi_B(x,y,t)}{\partial t} = &
\bigg[- \frac{m_r}{2 } \nabla^{2}_{\perp}+ V_B(x,y,t)+  \\ &\sum_{j=A,B}^{} \mathcal{U}_{Bj} |\psi_j(x, y, t)|^2 \bigg] \psi_B(x, y, t).
\end{split}
\end{equation}
In these expressions, $\nabla_{\perp} ^ 2 = \partial^2_x + \partial^2_y$ is the kinetic energy term, while $V_A(x,y)=(x^2 + y^2)/2$ and $V_B(x, y, t) = (x^2 + y^2)/2m_r + V_L$ 
denote the external potential of each species with mass ratio $m_r = m_A/m_B$. 
Moreover, $\mathcal{U}_{AA} = 2N_A \sqrt{2 \pi \lambda}a_{AA}/a_{osc}$ and $\mathcal{U}_{BB} = 2 N_B m_r \sqrt{2 \pi \lambda} a_{BB}/a_{osc}$ refer to the $A$ and $B$ 
species intraspecies interaction strengths respectively while the interspecies one corresponds to $\mathcal{U}_{AB} = \sqrt{2 \pi \lambda} N_A(m_A + m_B)a_{AB}/m_{B}a_{osc}$. 
Also, $N_{\sigma}$ with $\sigma=A,B$ denotes the particle number of each species which is taken to be $N_A =N_B= 10^4$ throughout this work.

\subsection{Ingredients of the numerical simulations}\label{Ingredients of the numerical methods} 

To study the dynamics of the binary bosonic system, we numerically solve Eqs. \eqref{3} and \eqref{4} using a split-time 
Crank-Nicolson method \cite{MURUGANANDAM2009, Ivana2012} adapted for binary condensates. 
In particular we obtain the initial state of the system, considering $\beta(t) = 1$ in Eq. \eqref{4}, by propagating Eqs. \eqref{3} and \eqref{4} 
in imaginary time, until the solution converges to the desired state. 
This state refers to either a singly quantized vortex  or the ground state of species A 
and a $\pi$ relative phase difference between the segments of species B. 
To introduce a vortex of charge $l$ at the center of the A species we apply the following transformation $\psi_{A} \rightarrow \psi'_{A} \exp[i l \tan^{-1}(\frac{y}{x})]$ 
to the corresponding initial guess wavefunction $\psi'_{A}$. 
Indeed this transformation, as we shall see below, results in a topological phase defect in $\psi_A$ around which arg($\psi_A$) winds by $2\pi l$. 
A similar topological phase imprinting method has been employed experimentally~\cite{Leanhardt2002} and theoretically~\cite{Bandyopadhyay2017,Kuopanportti2019}. 
Simultaneously, we impose the phase condition for the B species such that $\theta_B = \pi$ for $x \leq 0$ and $\theta_B = 0$ for $x> 0$ at the initial guess wavefunction. 
We remark that imprinting the phase only at the initial guess wavefunctions is adequate here since the spatial overlap between the two species is vanishing for our setup due to 
the presence of the potential barrier in species B. 
However, for spatially overlapping components one needs to imprint the corresponding phase at each time-instant of the imaginary time propagation in order to maintain the imprinted phase structure. 
Furthermore, the normalization of the $\sigma$ species wavefunction is ensured by utilizing the transformation $\psi_\sigma \rightarrow \frac{1}{\norm{\psi}} \psi_\sigma$ 
at every time-instant of the imaginary time propagation until the energy of the desired configuration 
is reached with a precision $10^{-8}$. 
With these solutions at hand as initial conditions, at $t = 0$, we study their evolution in real time up to $500$ ms utilizing Eqs. (\ref{3}) and Eq. (\ref{4}). 
The simulations are carried out in a square grid consisting of $400 \times 400$ grid points with a grid spacing $\Delta x = \Delta y = 0.05 $. 
The time step of integration $\Delta t$ is chosen to be $10^{-4}$. 
In this way we ensure the numerical convergence of our simulations.

\subsection{Observables}
  
To monitor the quench induced dynamics of the binary BEC, we utilize various observables.  
In order to investigate the dynamical formation of the vortices, we employ as a spatially resolved measure the vorticity 
of the $\sigma$ species namely 
\begin{equation}  
\Omega_{\sigma} = \nabla \times J_{\sigma},
\end{equation}
with $J_{\sigma} = \frac{i}{2m_{\sigma}}(\psi_{\sigma}^*\nabla\psi_{\sigma}- \psi_{\sigma}\nabla\psi_{\sigma}^*)$ being the velocity probability current. 
The overall vorticity of the system can be quantified via the integrated vorticity \cite{Ruben2008} 
\begin{equation}\label{6}
\mathcal{Z}_{\sigma} = \int \abs{\Omega_{\sigma}} dx dy.
\end{equation} 
We remark that, while $\Omega_{\sigma}$ offers a spatially resolved measure of the vorticity of each bosonic cloud, 
$\mathcal{Z}_{\sigma}$ reveals the overall \textit{vortical content} of the $\sigma$ species. 
To explore the angular momentum transfer between the different species we calculate the expectation value of the angular momentum of each species 
along the $z$ direction
\begin{equation}\label{5}
L^{\sigma}_z = -i \int\psi^*_{\sigma}(x\frac{\partial }{\partial y} - y\frac{\partial }{\partial x})\psi_{\sigma}  dx dy .
\end{equation}

Moreover, the total energy of the system reads 
\begin{equation}\label{en2}
 \begin{split}
  E(\psi_A, \psi_B) = & \int \bigg [ \frac{\abs{\nabla \psi_{A}}^2}{2} + V_{A}(x, y)\abs{\psi_{A}}^2 \\
  & + \frac{m_r\abs{\nabla \psi_{B}}^2}{2} + V_{B}(x, y)\abs{\psi_{B}}^2 \\
  & + \sum_{\sigma = A, B}^{}\frac{\mathcal{U}_{\sigma \sigma'} }{2} \abs{\psi_{\sigma'}}^2\abs{\psi_{\sigma}}^2  \bigg ].
 \end{split}
\end{equation}

To gain a deeper understanding of the different parts of the energy during the quench induced dynamics, it is appropriate to express 
the condensate wavefunction according to the Madelung transformation \cite{madelung1927quantentheorie}, namely 
$\psi_{\sigma}(x, y, t) = \sqrt{n_{\sigma}(x, y, t)}e^{i l_{\sigma} \theta_{\sigma}(x, y, t)}$, where $n_{\sigma}(x, y,t)$ and $\theta_{\sigma}(x, y,t)$ denote the density 
and the phase of the $\sigma$ species respectively and $l_{\sigma}$ is the $\sigma$ species vortex charge. 
Furthermore, we define $\sqrt{n_{\sigma}} {\bf u_{\sigma}}$
as the superfluid velocity weighted by the square root of the density, with $\bf u_{\sigma} = \nabla \theta_{\sigma}$.  
As a consequence, the total energy functional of Eq. (\ref{2}) can be decomposed as follows 
\begin{equation}
E = \sum_{\sigma =A, B}^{} \big [ E^{\rm kin}_{\sigma} + E^{\rm q}_{\sigma} + E^{\rm tr}_{\sigma} + E^{\rm int}_{\sigma} \big ] + E^{\rm mu}.\label{kinetic} 
\end{equation}
In this expression, $E^{\rm kin}_{A} = (1/2) \int \abs{\sqrt{n_{\sigma}}\bf u_{\sigma}}^2 dx dy$ and  $E^{\rm kin}_{B} = (m_r/2) \int \abs{\sqrt{n_{\sigma}}\bf u_{\sigma}}^2 dx dy$ 
is the kinetic energy of the A and the B species respectively.
Furthermore, $E^{\rm q}_{\sigma} = (1/2) \int (\nabla \sqrt{n_{\sigma}})^2 \dd x \dd y $ denotes the quantum pressure energy, 
$E^{\rm tr}_{\sigma} = (1/2) \int n_{\sigma} V_{\sigma}(x, y, t) \dd x \dd y$ is the potential (trap) energy, 
$E^{\rm int}_{\sigma} = (1/2) \int \mathcal{U}_{\sigma\sigma} n_{\sigma}^2 \dd x \dd y$ refers to the self-interaction energy and 
$E^{\rm mu} =  \int \mathcal{U}_{AB} n_A n_B \dd x \dd y$ is the mutual interaction energy between the different species. 
We are particularly interested in the kinetic energy part since it captures the very essence of the dynamics at different time scales. 
Indeed, $E^{\rm kin}_{\sigma} $ can be further decomposed into a compressible $E^{\rm kin,c}_{\sigma}$ and an incompressible $ E^{\rm kin, ic}_{\sigma}$ part, which are 
associated with the kinetic energy of the sound waves and the swirls, respectively \cite{Nore1997}. 
This decomposition is performed as follows. 
The velocity vector, $\sqrt{n_{\sigma}} {\bf u_{\sigma}}$ can be written as a sum over a solenoidal part, ${\bf u^{ic}_{\sigma}} $, and an irrotational part, ${\bf u^{c}_{\sigma}}$, 
namely,  $\sqrt{n_{\sigma}}{\bf u_{\sigma}} = \sqrt{n_{\sigma}} {\bf u^{c}_{\sigma}} + \sqrt{n_{\sigma}} {\bf u^{ic}_{\sigma}}$
such that $\div{\bf u^{ic}_{\sigma}} = 0$ and $\curl{\bf u^{c}} = 0$.  
We next define the scalar, $\Phi$, and vector, $A$, potential of the velocity field which satisfy the relations  
$\sqrt{n_{\sigma}} {\bf u^{ic}_{\sigma}} = \curl{{\bf A}}$ and $ \sqrt{n_{\sigma}}
  {\bf u^{c}_{\sigma}} = {\nabla} \Phi$ respectively. 
  Taking the divergence of the last expression we obtain the Poisson equation for the scalar field, i.e. $\laplacian{\Phi} = \div{\sqrt{n_{\sigma}} {\bf
    u_{\sigma}}}$. 
From this equation we can numerically determine the scalar potential \cite{Benyam2015} and consequently the corresponding field components ${\bf u^c_{\sigma}}$ 
and ${\bf u^{ic}_{\sigma}}$. 
Hence, the compressible and incompressible parts of the  species kinetic energy $A$ [$B$] species can be explicitly written as   
$E^{\rm kin, c}_{A} = 1/2 \int \abs{\sqrt{n_{A}} {\bf u^{c}_{A}}}^2 dx dy$ [$E^{\rm kin, c}_{B} = (m_r/2) \int \abs{\sqrt{n_{B}} {\bf u^{c}_{B}}}^2 dx dy$] and 
$E^{\rm kin, ic}_{A} = (1/2) \int \abs{\sqrt{n_{A}}{\bf u^{ic}_{A}}}^2 dx dy$ [$E^{\rm kin, ic}_{B} = (m_r/2) \int \abs{\sqrt{n_{B}}{\bf u^{ic}_{B}}}^2 dx dy$] respectively. 

\section{Dynamics of a slightly mass imbalanced mixture}\label{Equal_Mass} 

Let us focus here on the case of a small mass imbalance of the two species of the binary BEC, which in particular 
corresponds to the experimentally relevant mixture of $^{87}$Rb-$^{85}$Rb atomic species~\cite{Papp2008}. 
Moreover, the harmonic oscillator potential possesses a frequency, $\omega_A = \omega_B = 2\pi \times 30.83Hz$ and the 
typical anisotropy parameters $\alpha = 1$ and $\lambda = 40$ of the trapping which have been realized experimentally \cite{Vogels2001}. 
It is important to mention at this point that it might be difficult to achieve species selective potentials for a binary BEC 
composed of different hyperfine states of the same atomic species. 
Indeed, commonly the detuning responsible for optical trapping is much larger than the underlying hyperfine spltting. 
Such a difficulty has been rectified only very recently by employing the so-called ``tune-out'' approach~\cite{Bloch2018}. 
However, mixtures consisting of two distinct atomic species can be routinely trapped in species selective potentials exploiting both the ``tune-in'' and 
the ``tune-out'' optical trapping methods, see for details of the schemes Ref.~\cite{LeB2007}.
Hence, employing a slightly mass imbalanced mixture the advantage is two fold. 
First, it alleviates the possible experimental challenges to realize a species selective potential. 
Second, the dynamical behavior of the  issystem similar to the exactly mass balanced scenario. 
We shall also discuss how a significant mass imbalance between the species alters the quench-induced dynamics in section \ref{unequal_Mass}.

\subsection{Singly quantized vortex in species A} \label{Discussion1}
   
We start the discussion of our numerical results by investigating
the nonequilibrium dynamics of a 2D binary bosonic system when a vortex of unit charge is imprinted in species A 
and $\pi$ phase difference occurs between the segments of species B. 
Experimentally, a single BEC can be segmented into two slices by deforming a harmonic trap into a double-well potential utilizing 
a radiofrequency-induced \cite{Zobay2001, Jo2007} or an optical dipole potential \cite{Wang2005} along a certain spatial direction. 
Also, a well-defined relative phase difference between the two emergent slices of the condensate can be established using a single two 
photon coupling pulse \cite{Hall1998_measure}. 
A (quasi-one-dimensional) phase imprinting protocol along the lines of~\cite{Becker2008} can
have a similar effect. 
We consider the experimentally relevant intraspecies scattering lengths~\cite{Egorov2013,Papp2008}, $a_{AA} = 100.4a_0$ and $a_{BB} = 95.0a_0$, 
where $a_0$ denotes the Bohr radius. 
Also we assume an interspecies scattering length $a_{AB} = 60a_0$ and therefore the two species are in the miscible regime since 
$a_{AB}^2<a_{AA}a_{BB}$ is fulfilled \cite{Ao1998, Timmermans1998} .
After preparing the above-mentioned system in its initial state characterized by specific intra- and interspecies interactions, 
the dynamics is triggered by a sudden ramp down of the potential barrier of species B. 
The height and width of the optical potential barrier are chosen to be $U_0=20.0  \omega_x$ and $a=1.0a_{\rm osc}$ respectively.
This latter quench protocol allows the two initially separated segments of species B to collide and perform a counterflow dynamics. 
The postquench dynamics of the system can be categorized into three different time scales. 
Indeed, as we shall discuss below, at the very early stages of the dynamics quasi-1D nonlinear fringes known as 
dark stripes are formed. 
These dark stripes being subjected to the snake instability \cite{verma2017} break into several VAV pairs. 
Subsequently, these VAV pairs propagate within the condensate, interact with each other and eventually only a single VB soliton 
persists in the long-time dynamics of the system. 
Below we analyze the dynamics within each of the aforementioned time intervals in detail.

\begin{figure}[ht]
   \centering
   \includegraphics[width=0.5\textwidth]{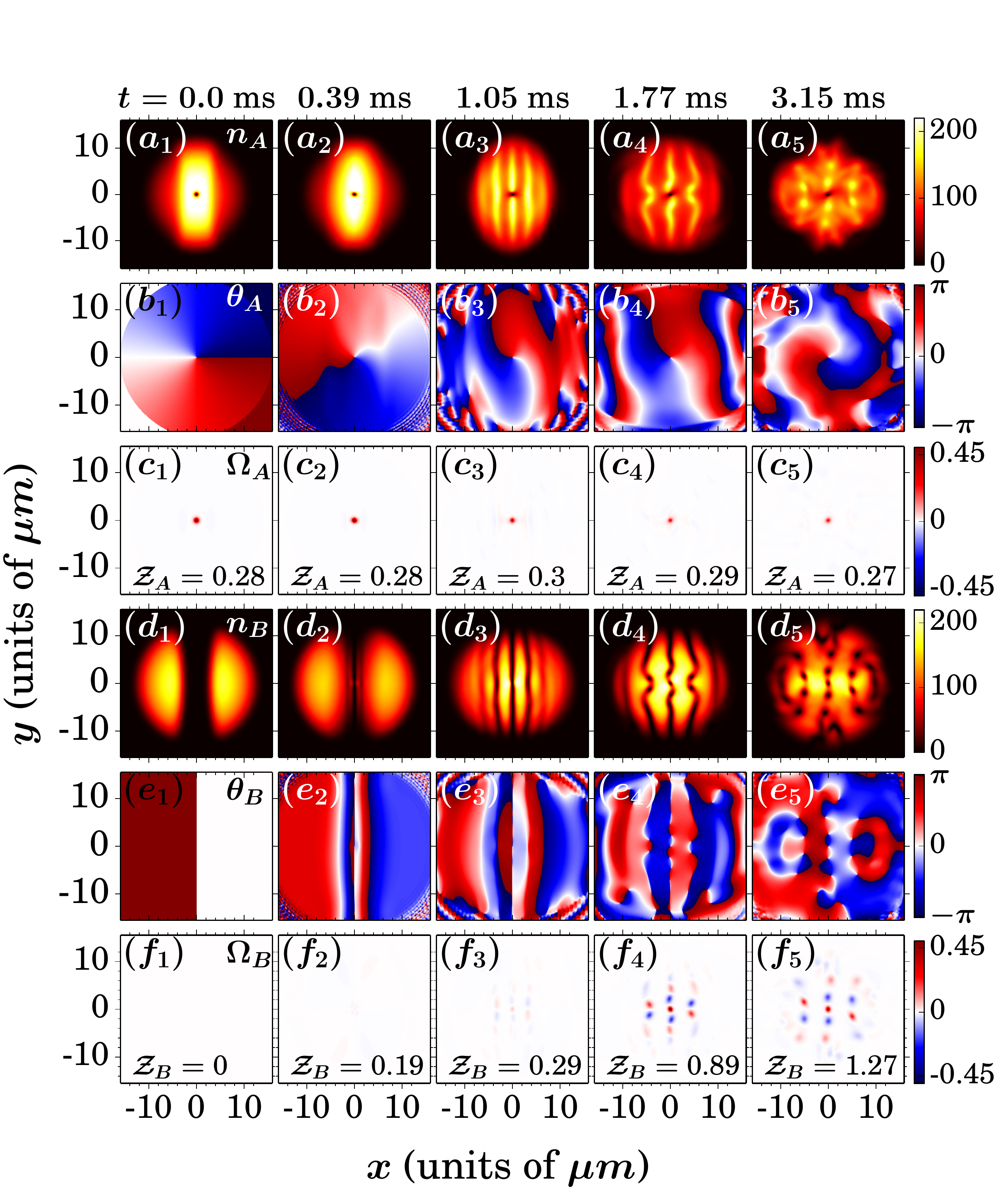}
   \caption{(Color online) Density profiles of ($a_1$)-($a_5$) species A and ($d_1$)-($d_5$) species B at various time instants (see legends) 
   in the very early stages of the dynamics. 
   The corresponding phase [vorticity] profiles of ($b_1$)-($b_5$) [($c_1$)-($c_5$)] species A and of ($e_1$)-($e_5$) [($f_1$)-($f_5$)] species B. 
   The values of the $\sigma$ species integrated vorticity $\mathcal{Z}_\sigma$ are also shown within each vorticity profile.  
   The binary BEC consisting of $N_A=N_B=10^4$ atoms is prepared in its ground state with a singly quantized vortex in species A and a $\pi$ phase difference in species B such 
   that $\theta_B(x) = \pi$ for $x < 0$ and $\theta_B (x) = 0$ for $x \geq 0$. 
   The intra- and interspecies scattering lengths are  $a_{AA} = 100.4a_0$, $a_{BB} = 95a_0$ and $a_{AB} = 60.0a_0$ respectively.}   
   \label{fig:1}
   \end{figure}

\subsubsection{Density evolution}\label{densities_1}

Figure \ref{fig:1} presents snapshots of the density, phase and vorticity profiles of both species at the early stages of the dynamics. 
We remark that due to the mutual interaction between the two species and the location of the potential barrier in species B, the initial density distribution of species A 
is elongated along the $y$-direction and in particular resides between the segments of species B [see Figs. \ref{fig:1} ($a_1$) and ($c_1$)]. 
As can be seen, the initial state of the system at $t=0$ ms, [Figs. \ref{fig:1} $(a_1), (b_1), (c_1), (d_1), (e_1), (f_1)$], breaks into dark stripes in species B  
[Figs. \ref{fig:1} ($d_2$)-($d_4$), ($e_2$)-($e_4$) and ($f_2$)-($f_4$)] and bright stripes
in species A  [Figs. \ref{fig:1}  ($a_2$)-($a_4$), ($b_2$)-($b_4$), ($c_2$)-($c_4$)] 
generated in the system directly after the merging of the two segments of species B. 
Subsequently, a disintegration of those stripes into vortices in species B [Figs. \ref{fig:1} $(d_5), (e_5), (f_5)$] and localized 
density humps in species A [Figs. \ref{fig:1} $(a_5)$] occurs. 

In particular, directly after turning off the potential barrier the two segments of species B smoothly expand towards each other, while 
species A expands along the $x$- direction. 
The merged segments, initially possessing a $\pi$- relative phase difference, resemble an excited mode of the harmonic potential at $t > 0$, see in particular 
a profile of the density in Fig. \ref{fig:1} ($d_2$) at $y=0$. 
As a result, a dark stripe of null density at the collision line accompanied by two lumps of higher atom density on its side are formed in species B [see Fig. \ref{fig:1} ($d_2$)].
At the same time, species A becomes more squeezed towards the collision line of species B leading to the formation of a bright stripe at the same place besides the trap center 
where the vortex is located [see Fig. \ref{1} ($a_2$)]. 
Subsequently, the localized density dips (in species A) and humps (formed in species B) split into a larger number of dark and bright stripes respectively.  
For instance, as it can be noticed from the density profiles of species A (B) at $t =1.05$ ms, multiple bright (dark) stripes occur  within
the condensate density. 
Focusing on species B, we observe that, due to the non-uniform density of the condensate (due to the presence of the external trap and associated curvature),
each dark soliton acquires a different velocity along the transverse $x$ direction. 
This leads to the bending of the soliton stripe and the resulting
undulations lead to its eventual breakup
into several VAV pairs \cite{Swartzlander1991, Toikka2013} as 
illustrated in Figs. \ref{fig:1} ($a_3$)-($a_5$) and ($d_3$)-($d_5$). 
Indeed, the formation of a chain of VAV pairs along the dark stripes, including a single vortex at the center of the middle stripe, 
can be noticed at $t =1.77$ ms and also even more pronouncedly
at $t=3.15$ ms. 
Note that a decreasing amplitude of the dark solitons implies an increase of its transverse velocity \cite{Tikhonenko1996}. 
Hence, due to the non-uniform density, the soliton stripes located close to the edge of the condensate bend faster than the ones residing in the middle 
[see also Figs. \ref{fig:1} ($d_3$) and ($d_4$)]. 
Moreover, as can be seen in Fig. \ref{fig:1} $(a_3)$, dark stripes created in species B are filled by the 
atoms of species A, thus the density of species A localizes at the same place.
This is natural as the former component acts as an effective potential
for the latter one. 
Therefore, the generation of vortices in species B is accompanied by the formation of localized density humps (characterized by no phase jump across them) 
in species A resembling this way VB structures.  
We remark that such density peaks on the top of the BEC background are also known as antidark solitons \cite{Kivshar1991}. 
Nevertheless, we will refer to these structures as VBs in what follows even when the bright component lies on top of a 
nontrivial background. 
Note that, bright solitons are bound states of the effective potential created by the vortices which stabilize the former. 
Otherwise, it would be impossible for the bright solitons to be sustained under the presence of repulsive interatomic interactions.

\begin{figure*}[ht]
\centering
\includegraphics[width=0.90\textwidth]{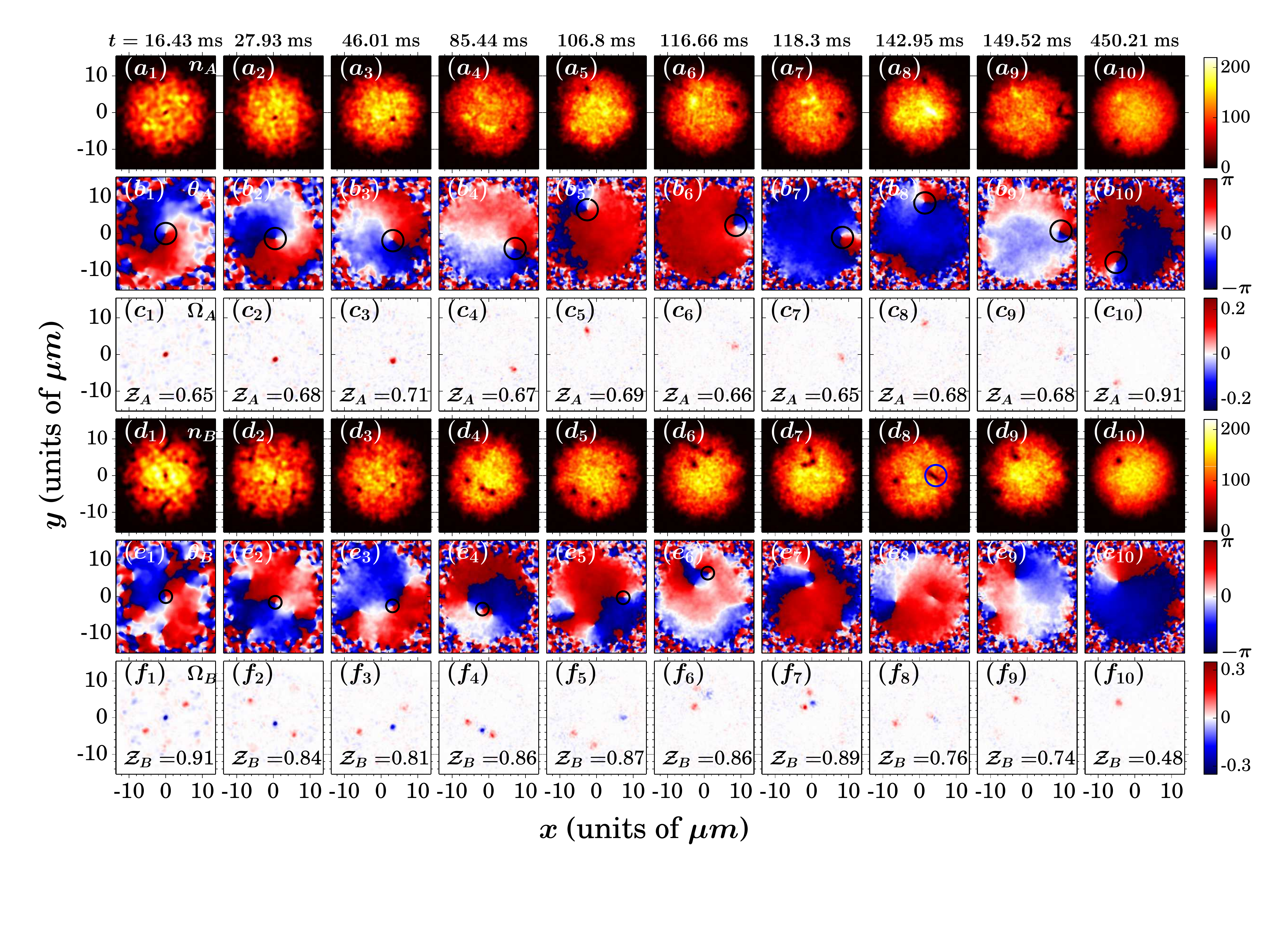}
\caption{(Color online) Density profiles of ($a_1$)-($a_{10}$) species A and ($d_1$)-($d_{10}$) species B at various time instants (see legends) 
in the long-time evolution of the system. 
The corresponding phase [vorticity] profiles of ($b_1$)-($b_{10}$) [($c_1$)-($c_{10}$)] species A and of ($e_1$)-($e_{10})$ [($f_1$)-($f_{10}$)] species B. 
The value of the $\sigma$ species integrated vorticity at each depicted time instant is also provided within the corresponding vorticity profiles. 
The black circles in the phase profiles mark the location of vortices. 
The remaining system parameters are the same as in Fig. \ref{fig:1}.} \label{fig:2}
\end{figure*} 

Having discussed the early stages of the dynamics, we next examine the longer time evolution of the system where a very rich dynamics 
of the generated VAV pairs takes place. 
Let us note in passing that the Gross-Pitaevskii equation has been extensively used to describe the long-time dynamics 
of similar setups \cite{Ruben2008, Tsatsos2016}. 
Also, possible interparticle correlation effects are expected to be supressed due to the large particle number considered herein. 
The long-time dynamics is summarized in Fig. \ref{fig:2}, which presents the density and vorticity profiles at selected time instants. 
After their production, the VAV pairs start to drift out of the condensate or they annihilate each other, see Fig. \ref{fig:1} ($d_5$) and Fig. \ref{fig:2} ($d_1$).  
The latter process reduces the number of vortices in species B. 
Indeed, inspecting Figs. \ref{fig:2} $(d_1)$, ($e_1$) and $(f_1)$, at $ t = 16.43$ ms, we observe that the VAV-V structure, which contains an antivortex at the center 
and two vortices placed on either side of the antivortex, is surrounded by several vortices [see Fig. \ref{fig:2} $(d_1)$]. 
Gradually, these latter randomly spaced vortices around the VAV-V structure vanish, and only the VAV-V remains 
[see Fig. \ref{fig:2} $(d_2)$ and $(f_2)$] in species B. 
In the course of the dynamics, both vortices approach the antivortex several times and scatter [e.g. see Figs. \ref{fig:2} ($d_2$)-($d_5$) and ($f_2$)-($f_5$)].  
After such a scattering event the VAV-V moves around the condensate keeping its linear arrangement intact [e.g. see Fig. \ref{fig:2} ($f_2$)-($f_4$)] 
or it precesses around the trap center where the linear arrangement of its constituents is distorted to a triangular one [Fig. \ref{fig:2}($f_3$)-($f_5$)] \cite{Seman2010}. 
Also, during the dynamics, one of the outermost vortices of the VAV-V reaches the surface of the cloud and excites surface modes. 
In turn, these surface modes can isolate the vortex (breaking in this
way the VAV-V) at the edge of the cloud [see Fig. \ref{fig:2} $(d_6)$ and $(d_7)$]. 
At $t=142.95$ ms, as a result of the above breakup
we observe that a structure is formed by the coalescence of the VAV pair near the edge of the cloud, see also the blue circle in Fig. \ref{fig:2} $(d_8)$.  
Such a structure, is sometimes referred to
as vortexonium \cite{Simula2016}, and it is hard to discern individual vortex phases although it remains a spatially localized bound state.
More appropriately, this structure is known as the Jones-Roberts (JR)
soliton (see, e.g., the recent discussion in~\cite{Chiron_2018})
and it has been recently considered also experimentally~\cite{bongs18}.
As can be seen in Fig. \ref{fig:2} ($d_9$), the structure
propagates towards the trap center and finally drifts out of the condensate leaving behind only a single vortex (from the original VAV-V structure). 

On the other hand, the bright solitons generated in species A are located at the core of the vortices of species B. 
Recall that the existence of the bright solitons in species A is caused by the occurrence of vortices in species B. 
These bright solitons, being density humps on top of the BEC background of species A, certainly contribute to the inhomogeneity of 
the density of species A. 
Recall that a vortex experiences a density gradient from the condensate \cite{Bolda1998, Jackson1999} which in turn 
creates a velocity field affecting its motion. 
If this density gradient is symmetrically distributed with respect to the position of the vortex, then the net velocity field acting on the vortex is zero and the vortex remains  
stationary in its original position. 
A close inspection of Figs. \ref{fig:2} ($a_2$)-($a_3$) reveals that the bright soliton structures building upon species A are not 
symmetrically distributed with respect to the trap center. 
Therefore, the vortex in species A experiences a net velocity field, towards the region of lower condensate density, which eventually displaces it from the 
trap center towards the periphery of the cloud. 
Subsequently, the displaced vortex precesses around the trap center \cite{Anderson2000}, see Figs. \ref{fig:2} ($a_4$)-($a_{10}$), 
and persists in the long-time dynamics (see the discussion bellow).

\subsubsection{Time evolution of the integrated vorticity}\label{vorticity_and_kinetic_conf1} 

The dynamics of the system can further be understood by inspecting the time evolution of the integrated vorticity and of the different parts of the kinetic 
energy [see Eq. (\ref{kinetic})] of the system.  
Recall that the integrated vorticity defined in Eq. (\ref{6}) is a quantitative measure of the effects arising due to the \textit{vortical content} of 
each species. 
In particular, if the $\sigma$ species does not (initially)
contain any net vorticity, the area integral of $\Omega_{\sigma}$ without the modulus is zero owing to the cancellation 
between the regions of opposite vorticity. 
Clearly, this is not the case with the integrated vorticity, $\mathcal{Z}_{\sigma}$, which accounts for the overall vorticity present in the $\sigma$ species. 
As it can be seen from the inset of Fig. \ref{fig:3} $(a)$, the integrated vorticity  of species A is initially non-zero, i.e.  $\mathcal{Z}_A \ne 0$, due to the presence 
of the singly quantized vortex in species A.  
However, $\mathcal{Z}_B = 0$, since at $t=0$ no vortex is present in species B. 
Moreover, the integrated vorticity of both species increases [Fig. \ref{fig:3} $(a)$] within the time interval $0$ ms $<t<2$ ms since the collision of the segments of species B 
gives rise to a rapid generation of vortices and antivortices [e.g. see also Figs. \ref{fig:1} ($d_3$) and ($d_5$)]. Notice that this increase is associated
with the absolute value in the definition of the relevant quantity,
indicating that this diagnostic will increase even though there is no
net vorticity gain within the dynamics. 
Then, $\mathcal{Z}_B$ decreases till $t = 16$ ms, which suggests that the number of vortices in species B reduces, until the VAV-V structure is formed 
[see also Fig. \ref{fig:1} ($d_5$) and Fig. \ref{fig:2} ($d_1$)]. 
Within the time period between the formation of the VAV-V structure and the final
single vortex stage, $\mathcal{Z}_B$ shows a highly fluctuating behavior around a mean value. 
On the other hand, $\mathcal{Z}_A$ increases upon the removal of the barrier but the overall vorticity of species B remains always larger than that of species A 
throughout the time evolution. 
After a single vortex finally persists in species B, $\mathcal{Z}_A$ and $\mathcal{Z}_B$ tend to approach each other. 

\subsubsection{Kinetic energy contributions}\label{kinetic_conf1}

To shed further light onto the vortex dynamics of the system, we explore the time evolution of the different parts of the kinetic energy of each species. 
Note that, the incompressible kinetic energy, $E^{\rm kin, ic}_{\sigma}$, reveals the presence of vorticity in the $\sigma$ species, whereas the compressible kinetic 
energy, $E^{\rm kin, c}_{\sigma}$, is associated with the presence of acoustic waves. 
Let us first concentrate on the kinetic energy of species B. 
At $t=0$, there is no contribution to the energy from its kinetic part [see the right inset of Fig. \ref{fig:3} (b)]. 
However, shortly after the quench, $0$ ms $< t < 0.6$ ms, a rapid growth of $E^{\rm kin, c}_B$ is observed indicating the generation of acoustic waves. 
In fact, the undulations of the dark soliton stripes contribute to the acoustic waves that lead to the generation of vorticity in the system. 
For later evolution times, $0.6$ ms $< t < 2.0$ ms the generation of vortices causes the increasing behavior of $E^{\rm kin, ic}_B$, while 
$E^{\rm kin, c}_B$ decreases.  
Indeed, $E^{\rm kin, c}_B$ acquires a minimum value at $t=2$ ms, when $E^{\rm kin, ic}_B$ is maximized. 
This suggests that the generation of vortices into the system within this time interval might be attributed to the conversion of the 
acoustic energy into the swirling energy. 
Subsequently, as most of the vortices drift out of the condensate or annihilate each other, and in the process emit acoustic 
waves \cite{Leadbeater2001}, a steep decrease of $E^{\rm kin, ic}_B$ accompanied by an increase of $E^{ \rm kin, c}_B$ occurs until $t = 16$ ms, 
where the above discussed VAV-V is formed [see also Fig. \ref{fig:3} (b)]. 
The VAV-V structure gradually converts its energy into acoustic energy within the time interval $16$ ms $\leq t \leq 149$ ms, and then finally a pair of
vortices is ejected into the background leaving behind a single vortex.

This final stage involving a single vortex can be noticed from the drop of $E^{\rm kin, ic}_A$ which afterwards remains almost stationary in time. 
By virtue of the initial vortex configuration, the kinetic energy of species A is dominated by its incompressible counterpart, $E^{\rm kin, ic}_A$.
Although species A is confined in a rotationally symmetric harmonic trap, due to the interspecies interaction its density is compressed to the region 
where the potential barrier of species B is located, i.e., species B
forms an effective potential which further confines species A. 
However, due to the interspecies interaction the merging process occurring in species B also affects the density of species A. 
In particular, as species A expands after the removal of the barrier, acoustic waves are also 
generated in species A, resulting in the rise of $E^{\rm kin, c}_A$ [see Fig.\ref{fig:3} (e)].  
Overall, $E^{\rm kin, ic}_A$ decreases while $E^{\rm kin, c}_A$ increases during the time evolution. 
Moreover, all components of the kinetic energy remain almost stationary for $ t \geq 330$ ms, when the vortical patterns in each component have essentially settled. 
However, fluctuations around their mean values reveal the occurrence of energy exchange among the compressible and the incompressible parts of the velocity field. 
It is also worthwhile to note that $E^{\rm kin, ic}_A$, possessing initially the dominant contribution of the kinetic energy, for $t \geq 50$ ms becomes 
the smallest contributor to the total kinetic energy of the system. 
This raises the very important question on how vorticity of species A decays over time, and whether it is transferred to species B. 

\subsubsection{Dynamics of the angular momentum}\label{angular_momentum}

\begin{figure}[ht]
\centering
\includegraphics[width=0.4\textwidth]{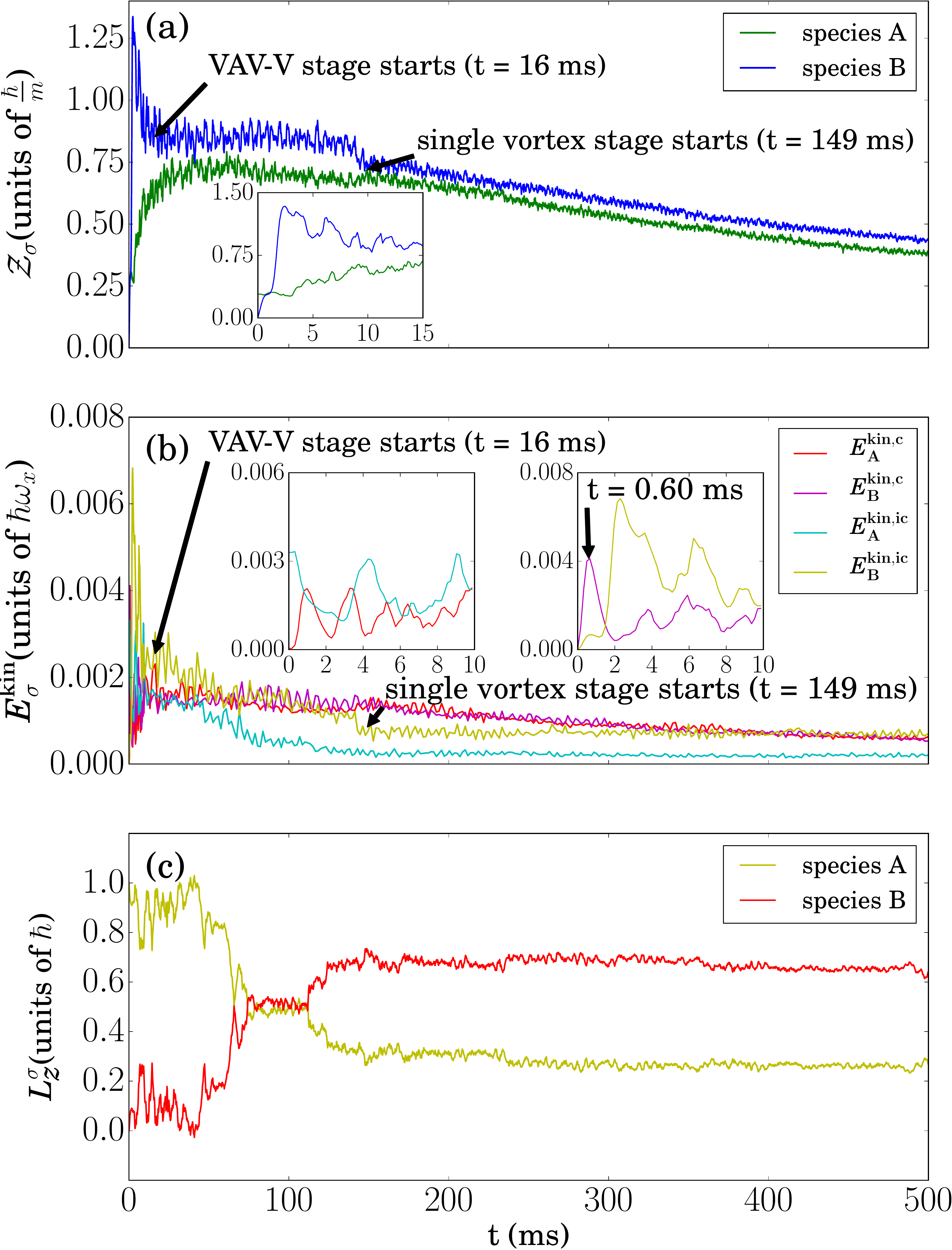} 
\caption{(Color online) (a) Time evolution of the integrated vorticity $\mathcal{Z}_{\sigma}(t)$ of each species. 
Inset shows $\mathcal{Z}_{\sigma}(t)$ within the time interval $0$ ms $<t<10$ ms. 
(b) Compressible, $E^{\rm kin, c}_{\sigma}$, and incompressible, $E^{\rm kin, ic}_{\sigma}$, parts of the kinetic energy of the $\sigma$ species (see legend) 
during the evolution. 
Insets present $E^{\rm kin, c}_A$ and $E^{\rm kin, ic}_A$  (left panel) and $E^{\rm kin, c}_B$ and $E^{\rm kin, ic}_B$ (right panel) at the 
initial stages of the dynamics. 
The black arrows in (a), (b) indicate the time instants that the VAV stage and the singly quantized vortex stage start in species B. 
(c) Dynamics of the angular momentum, $L^{\sigma}_z(t)$, of the $\sigma=A,B$ species along the $z$ direction. 
The remaining system parameters are the same as in Fig. \ref{fig:1}.}\label{fig:3}
\end{figure}

To further unravel the dynamics of the vorticity of both species, we resort to the expectation value of the angular momentum of each species, $L^{\sigma}_z(t)$ [see also Eq. (\ref{5})] 
along the $z$- direction, as shown in Fig. \ref{fig:3} (c).   
Evidently, at $t=0$ ms the average angular momentum per particle of species A is unity, while for species B is zero, given their respective vorticities. 
Remarkably enough, the angular momentum of each species is not conserved after the removal of the barrier. 
A rapid angular momentum exchange between the two species takes place throughout the time evolution and a net angular momentum is gained by species B 
resulting eventually in $L^{A}_z \le L^{B}_z$. 
Moreover, both species acquire a non integer angular momentum per particle. 
We remark here that the merging of the two segments in species 	B actually changes the rotational properties of the system by
leading it to acquire vortices in the periphery of the cloud and
thus to rotate in a nearly rigid-body rotation as a result~\cite{White2016, white2017}. 
It is also important to stress that the above-described angular momentum transfer among the different species depends on the initially imprinted relative 
phase difference between the segments of species B. 
For instance, herein, we found that a net angular momentum transfer is possible when a $\pi$ relative phase difference exists initially between the segments 
of species B. 
However, imprinting a zero phase difference between the segments, no such transfer process has been found (see also Appendix.~\ref{Derivation} 
for a detailed investigation of the dependence of the angular momentum transfer on the above-mentioned phase difference is certainly of interest but lies beyond the 
scope of the present work. 

To understand the above-described transfer of angular momentum between the species, we next derive the corresponding time evolution of the 
angular momentum of each species. 
Indeed it can be shown (see also Appendix B) that  
\begin{equation}\label{9}
  \begin{split}
      	\frac{\dd L^{A}_{z}}{\dd t} = & -\int  \abs{\psi_A(x, y)}^2(x\frac{\partial }{\partial y} - y\frac{\partial }{\partial x} )V_A(x,y) dx dy\\
      	& -\int\abs{\psi_A(x, y)}^2(x\frac{\partial }{\partial y} - y\frac{\partial }{\partial x} ) \mathcal{N}_A(x,y)dx dy
   \end{split}	   
\end{equation}
and 
\begin{equation}\label{10}
 \begin{split}
  \frac{\dd L^{B}_{z}}{\dd t} = & -\int\abs{\psi_B(x, y)}^2(x\frac{\partial }{\partial y} - y\frac{\partial }{\partial x} )V_B(x,y) dx dy \\
   & -\int\abs{\psi_B(x, y)}^2(x\frac{\partial }{\partial y} - y\frac{\partial }{\partial x} ) \mathcal{N}_B(x,y) dx dy.
  \end{split}	   
\end{equation}
In these expressions, $V_A$ and $V_B$ denote the external trapping potentials of species A and species B respectively [see also Eqs. (\ref{3}) and (\ref{4})]. 
Furthermore, $\mathcal{N}_A(x, y)$ = $\mathcal{U}_{AA}\abs{\psi_A(x,y)}^2 + \mathcal{U}_{AB}\abs{\psi_B(x,y)}^2$ and 
$\mathcal{N}_B(x,y) = \mathcal{U}_{BB}\abs{\psi_B(x,y)}^2 + \mathcal{U}_{AB}\abs{\psi_A(x,y)}^2$ 
are the corresponding nonlinear interaction energy terms. 
Most importantly, each of the Eqs. (\ref{9}) and (\ref{10}) can be divided into two parts. 
The first one includes the angular momentum operator and the external potential. 
This term vanishes for a spatially isotropic external potential and yields a non-zero contribution which becomes 
more significant for a larger spatial anisotropy. 
Since in our case we consider a sudden quench of the potential barrier then at $t>0$ the post-quench 
external potential is indeed isotropic and therefore this term vanishes. 
The second part of the Eqs. (\ref{9}) and (\ref{10}) captures the contribution from the nonlinear interaction term. 
In particular, it indicates that if the nonlinear interaction term is spatially isotropic then its contribution is zero but not if 
it is anisotropic. 
For our setup the densities of both species after the quench are anisotropic [see Figs. \ref{fig:1} ($a_2$) and ($d_2$)] and therefore this 
term is responsible for the observed angular momentum transfer. 
We have indeed verified the above-mentioned argumentation also within our simulations (results not shown here for brevity). 

\subsubsection{Effect of the interspecies interactions}

After having explained the mechanism behind the angular momentum transfer we next discuss how this transfer is affected by the interspecies interaction strength. 
More specifically, we are interested in investigating how the degree of miscibility impacts the angular momentum transfer. 
Recall that homogeneous binary condensates are termed miscible when $a^2_{AB} \leq a_{AA}a_{BB}$ is fulfilled, and immiscible when 
$a^2_{AB} \geq a_{AA}a_{BB}$ holds \cite{Ao1998, Timmermans1998}. 
However, under the presence of an external confinement this condition is modified due to the inhomogeneity of 
the density profiles \cite{navarro,Wen2012,bisset}. 
A well-known measure of the degree of the phase separation, namely the degree of miscibility or immiscibility, is the overlap 
integral between the two species \cite{Mistakidis_2018}
\begin{equation}\label{7}
\Lambda =
\frac{
	\biggr[
	\displaystyle\int\hspace{-0.2cm}\displaystyle\int{\rm d}x
	\hspace{0.05cm}{\rm d}y\hspace{0.05cm}n_{A}(x, y)
	\hspace{0.05cm}n_{B}(x,y)
	\biggr]^{2}
}
{
	\biggr[
	\displaystyle\int\hspace{-0.2cm}\displaystyle\int{\rm d}x
	\hspace{0.05cm}{\rm d}y\hspace{0.05cm}n_{A}^{2}(x, y)
	\biggr]
	\biggr[
	\displaystyle\int\hspace{-0.2cm}\displaystyle\int{\rm d}x
	\hspace{0.05cm}{\rm d}y\hspace{0.05cm} n_{B}^{2}(x, y)
	\biggr] 
}.
\end{equation} 
Here, $\Lambda(x, y, t) = 0$ and $\Lambda(x, y, t) = 1$ indicate zero
(i.e., phase separation)
and complete (i.e., phase mixing)
spatial overlap between the two species, respectively. 

\begin{figure}[ht]
	\centering
	\includegraphics[width=0.45\textwidth]{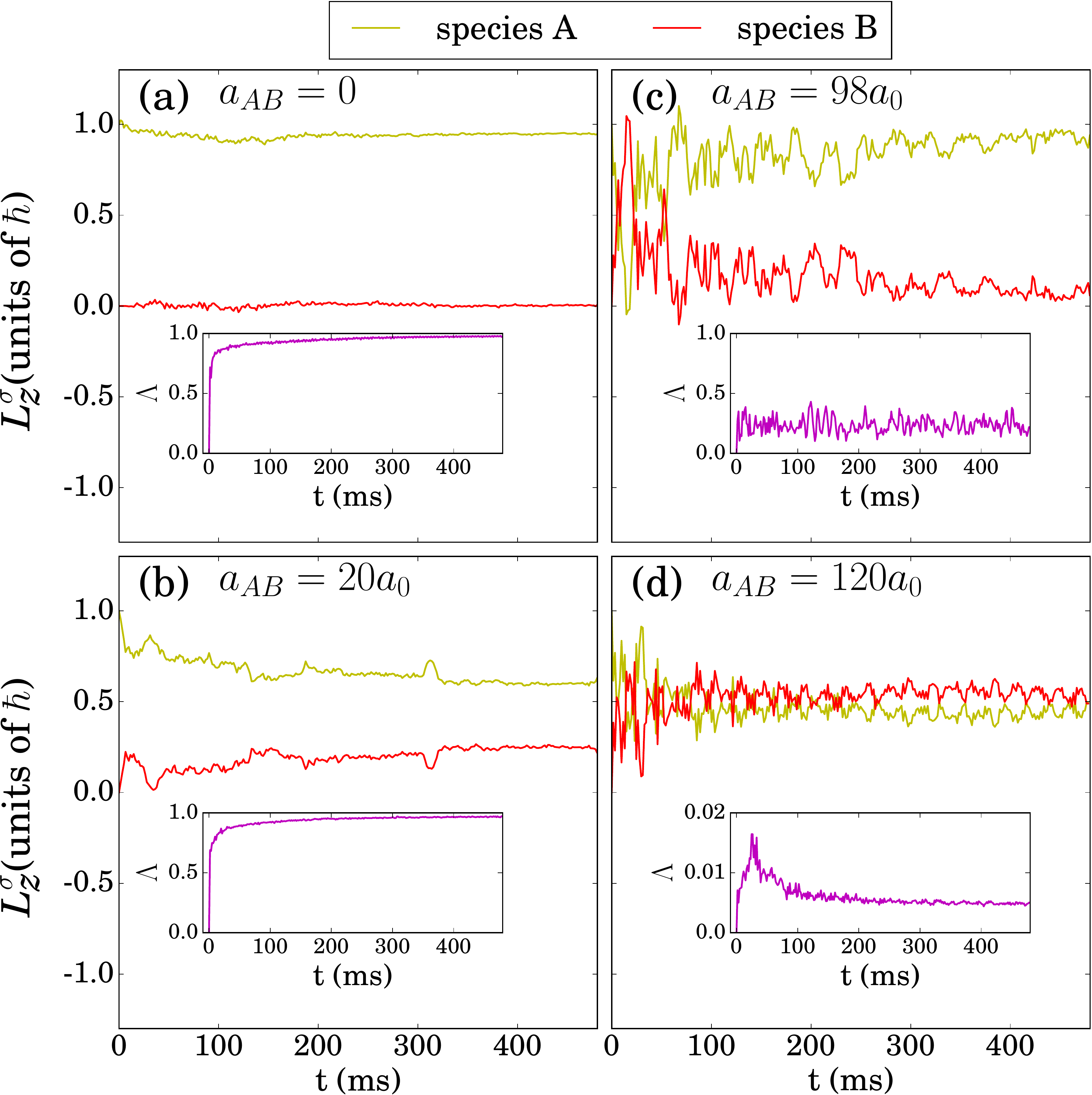}
	\caption{(Color online) Time evolution of the angular momentum, $L^{\sigma}_z(t)$, of the $\sigma$ species for (a) $a_{AB} = 0$, (b) $a_{AB} = 20 a_{0}$, 
		(c) $a_{AB} = 98 a_{0}$ and (d) $a_{AB} = 120 a_{0}$. 
		The insets show the evolution of the corresponding overlap integral. 
		The other system parameters are the same as in Fig. \ref{fig:2}. 
		The initial state of the mixture corresponds to an immiscible phase for all interspecies interactions due to the 
		presence of the potential barrier in species B.}\label{fig:4}
\end{figure}

Figure \ref{fig:4} presents $L^A_z(t)$ [$L^B_z(t)$] of species A [species B] for various interspecies interaction strengths. 
$\Lambda(t)$ is also illustrated in the corresponding insets. 
Note that, due to the presence of the barrier, $\Lambda(t)$ is zero initially
($t = 0$) for each interspecies interaction. 
For $a_{AB} = 0$, $L^{A}_z(t)$ and $L^{B}_z(t)$ evolve independently keeping their initial angular momentum almost intact [see Fig. \ref{fig:4}(a)]. 
This suggests that no angular momentum is transferred from species A to species B, while $\Lambda(t) \rightarrow 1$ rapidly
[see the inset of Fig. \ref{fig:4} (a)]. 
However, as $a_{AB}$ is increased to $20 a_{0}$, a small portion of angular momentum is transferred to species B, yet the angular momentum of species A 
remains always higher than that of species B.  
The corresponding overlap integral $\Lambda (t)$ rapidly approaches unity indicating a miscible behavior between the species [see the inset of Fig. \ref{fig:4} (b)]. 
In contrast to the above a notable angular momentum transfer between the species occurs when a sufficiently large interspecies interaction is introduced. 
Indeed, for $a_{AB} = 60a_{0}$, we observe that the angular momentum of species B exceeds that of species A [see Fig. \ref{fig:3} (c)]. 
This situation drastically changes at
the border between the miscible to the immiscible regime,
i.e.,  $a_{AB} = 98a_{0}$. 
In this  case species A transfers all of its angular momentum to species B just after the collision of the two segments. 
But shortly after the above-mentioned exchange species B transfers again its angular momentum back to species A, and both species feature 
an angular momentum exchange, until species A (approximately)
acquires its initial angular momentum [see Fig. \ref{fig:4} (c)]. 
Here, $\Lambda (t)$ shows small fluctuations in time around 0.25 as illustrated in the inset of Fig. \ref{fig:4} (c). 
Entering the phase separated regime with $a_{AB} = 120a_0$, the two species are initially spatially separated and they possess 
different angular momentum [see Fig. \ref{fig:4} (d)]. 
After the quench $t>0$ we observe an angular momentum transfer between the species and eventually at longer evolution times, $t>90$ ms, $L^{B}_z(t)$ becomes 
slightly larger than $L^{A}_z(t)$. 
This difference between $L^{A}_z(t)$ and $L^{B}_z(t)$ results in a nonzero overlap integral even in the immiscible regime [see the inset of Fig. \ref{fig:4} (d)]. 
Note here that the density of species A resides in between the segments of species B and therefore if $L^{A}_z(t)\neq L^{B}_z(t)$ the two species favor a spatial overlap between them. 
We remark that since $L^{A}_z(t)$ and $L^{B}_z(t)$ are finite in the long-time dynamics of the system, a single vortex builds upon each of the species and remains almost 
stationary for $t>100$ ms (results not shown for brevity).

\section{Dynamics of a slighlty mass balanced mixure with zero vorticity in species A}\label{Discussion2}

\begin{figure}[ht]
	\centering
	\includegraphics[width=0.50\textwidth]{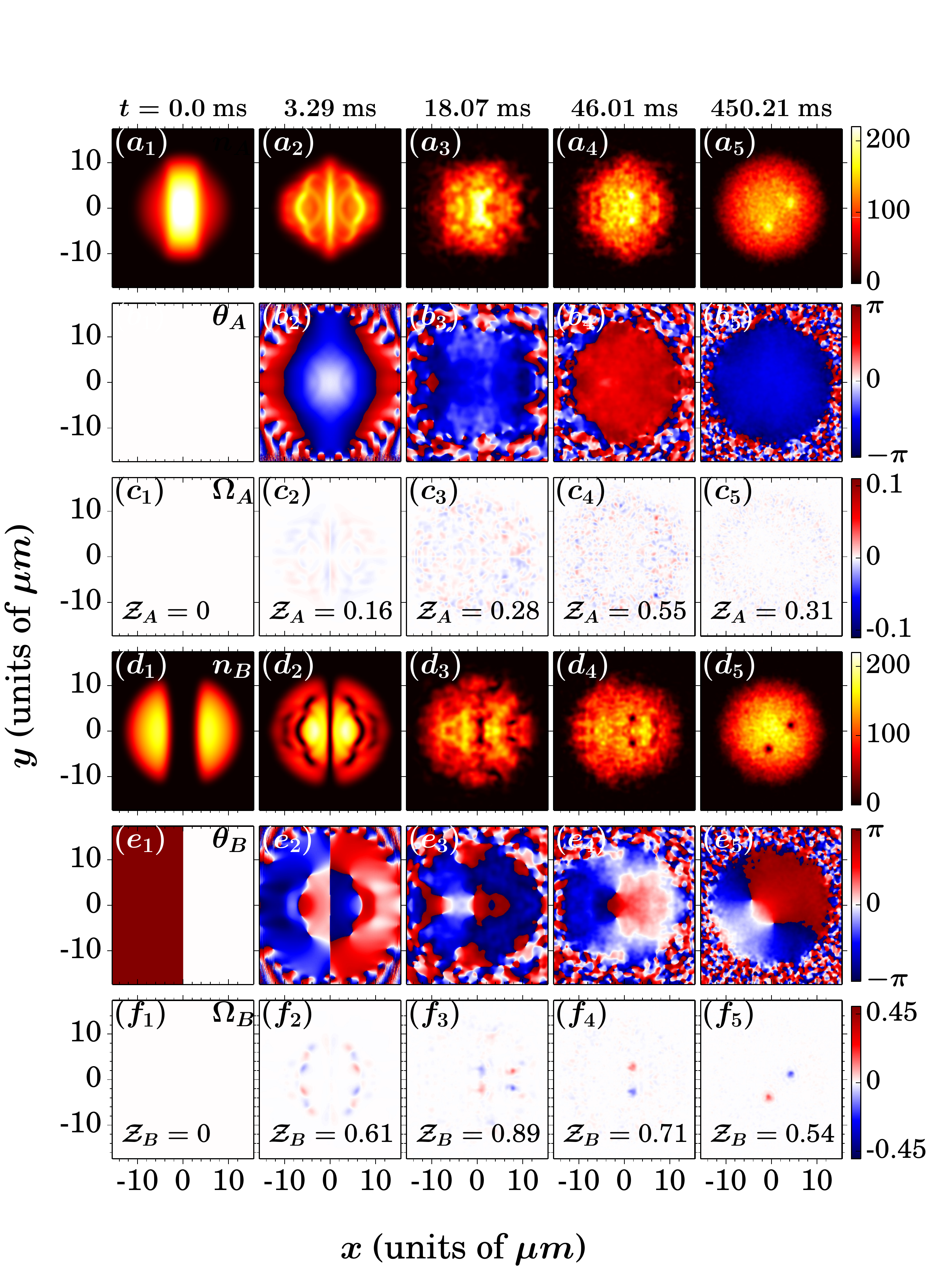}
	\caption{(Color online) Density profiles of ($a_1$)-($a_5$) species A and ($d_1$)-($d_5$) species B at selected time instants of the evolution. 
			Shown are also the corresponding phase [vorticity] profiles of ($b_1$)-($b_5$) [($c_1$)-($c_5$)] species A and ($e_1$)-($e_5$) [($f_1$)-($f_5$)] 
			species B. 
			The value of the $\sigma$ species integrated vorticity is provided within the corresponding vorticity profiles.
			The binary BEC consists of $N_A=N_B=10^4$ bosons and it is prepared in its ground state with no vortex in species A and a phase 
			difference in species B such that $\theta_B(x) = \pi$ for $x < 0$ and $\theta_B (x) = 0$ for $x \geq 0$.  
			The intraspecies scattering lengths are $a_{AA} = 100.4a_0$ and $a_{BB} = 95a_0$, while the interspecies scattering length 
			is $a_{AB} = 60.0a_0$ }. \label{fig:6}
\end{figure}

Here we analyze the emergent nonequilibrium dynamics of the $^{87}$Rb-$^{85}$Rb system when the initial configuration of species B contains a relative phase difference of $\pi$
between its segments, but species A possesses zero vorticity.  
Recall that species A is trapped in an isotropic 2D harmonic oscillator potential while species B is confined in an anisotropic trap consisting of a harmonic oscillator of equal frequencies
in both directions and a potential barrier in the $y$ direction. 
With this configuration as initial state of our system, we again suddenly ramp down the barrier at $t=0$ ms, and let the system to evolve till $t=500$ ms. 
The density, phase and vorticity profiles of both species at various time instants are presented in Fig. \ref{fig:6}. 
As in the previous section, we observe that after the quench the collision between the segments of species B generates dark and bright stripes 
in species B and species A respectively, see also Figs. \ref{fig:6} ($a_2$)-($d_2$). 
Subsequently, the dark and bright stripes break into VAV pairs [Fig. \ref{fig:6} ($d_4$)-($d_5$)] and bright solitons [Fig. \ref{fig:6} ($a_4$)-($a_5$)] respectively 
through the manifestation of the snake instability [see Fig. \ref{fig:6} ($d_2$)]. 
For long evolution times the aforementioned structures via their interactions
eventually result in vortex-bright-soliton entities [Fig. \ref{fig:6} ($a_5$), ($d_5$)].   
It is also important to note that the VAV pairs building upon species B, in the process of interacting and annihilating each other, produce two VAV dipoles.  
For instance, inspecting Figs. \ref{fig:6} ($d_3$), ($e_3$) and ($f_3$) we observe that one VAV pair resides at the trap center of the condensate, 
and another pair at the edges. 
However, this structure is not stable, since one vortex pair moves out of the condensate [see Fig. \ref{fig:6}($d_4$), ($e_4$) and $(f_4)$], resulting 
in a vortex dipole in species B and a pair of bright-solitons in species A. 

The VAV pair formed close to the trap center creates an effective potential localizing
bright-solitons of species A therein. 
We remark that a VAV pair with long lifetime has already been detected in numerous recent
experiments \cite{Neely2010, Middelkamp2011, Kwon2015},
where they are generated through a diverse set of experimental techniques
(including dragging of a repulsive light barrier or quenching through
the BEC transition). 
Recall that with the same initial phase configuration, species B produces
a single vortex, when species A carries initially one unit vortex charge. 
Hence, given the relative $\pi$ phase difference between the segments of species B, it is the initial phase configuration of species A which plays a crucial
role towards the generation of such topological defects (and especially so a net topological charge) in species B.

\begin{figure}[ht]
	\centering
	\includegraphics[width=0.42\textwidth]{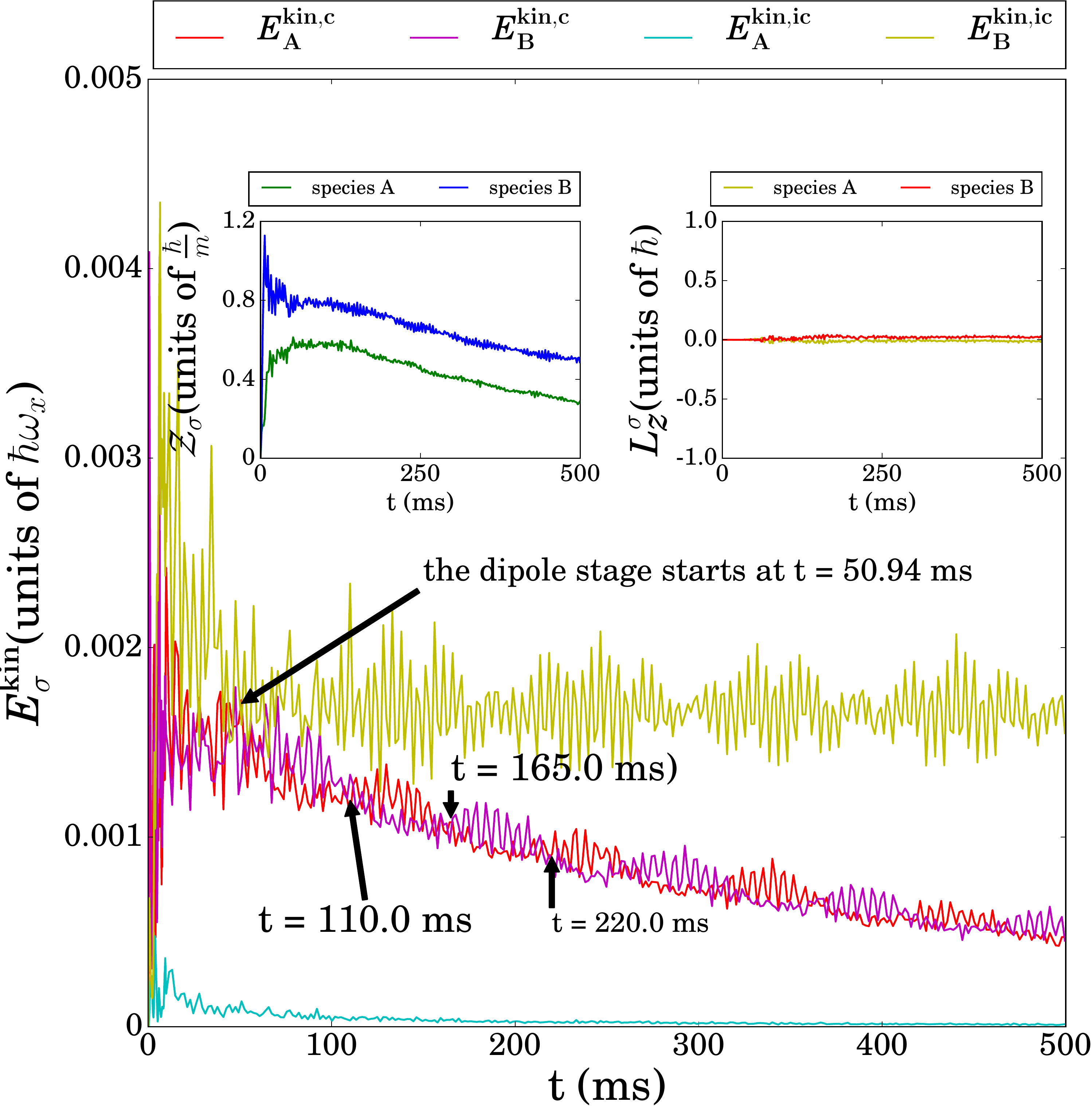}
	\caption{(Color online) Time evolution of the various parts of the kinetic energy (see legends). 
		Inset shows the integrated vorticity (left panel) and the angular momentum (right panel) during the dynamics. 
		The arrow indicates the time instant at which the dipole is created in species B.  
		The remaining system parameters are the same as in Fig. \ref{fig:6}}\label{fig:7}
\end{figure}

\begin{figure}[ht]
	\centering
	\includegraphics[width=0.48\textwidth]{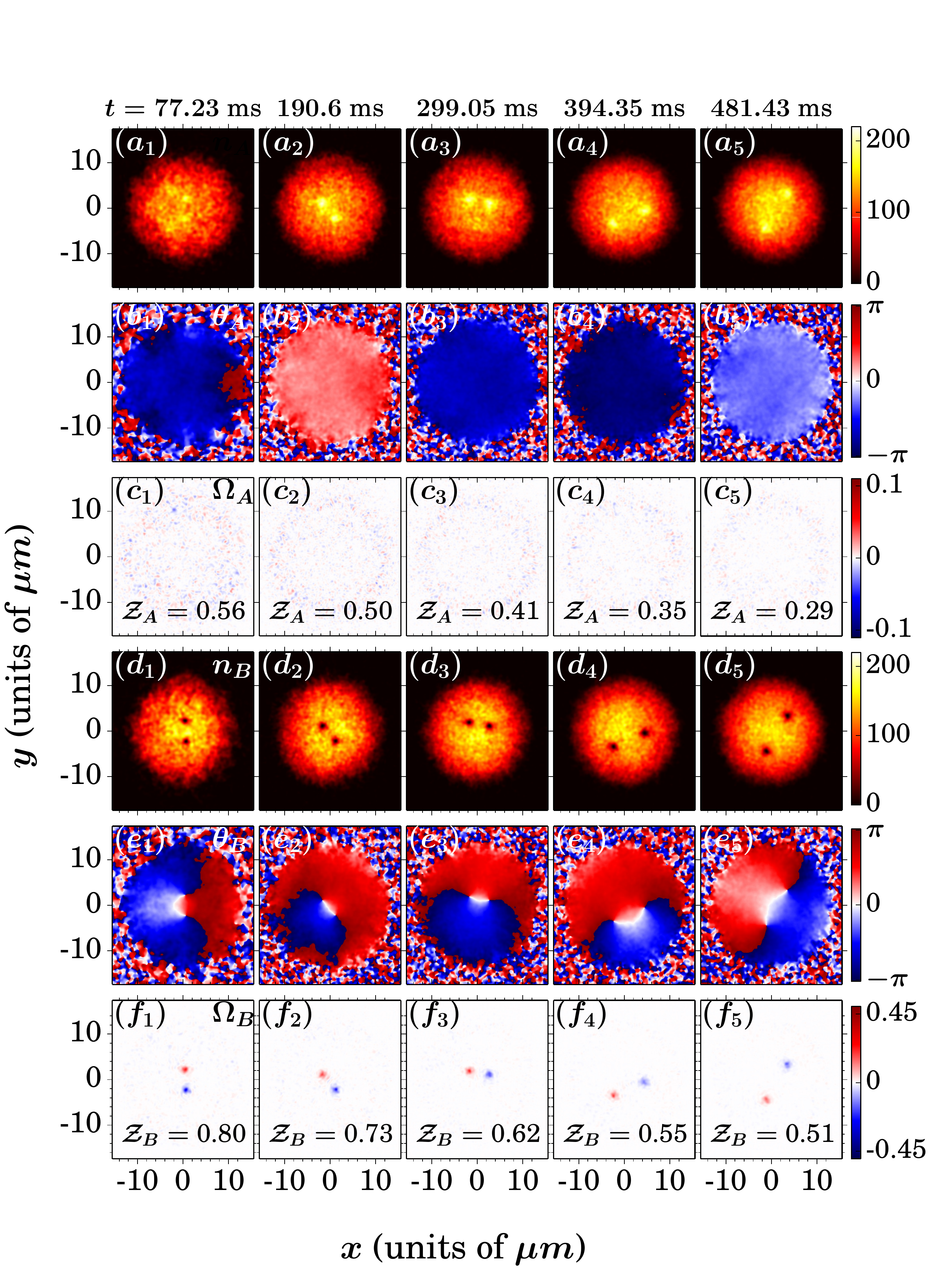}
	\caption{(Color online) Density profiles of ($a_1$)-($a_5$) species A and ($d_1$)-($d_5$) species B at long evolution times. 
			The respective phase [vorticity] profiles of ($b_1$)-($b_5$) [($c_1$)-($c_5$)] species A and ($e_1$)-($e_5$) [($f_1$)-($f_5$)] 
			species B are also presented in order to illustrate the dynamics of the vortex dipole. 
			The integrated vorticity of each species at the depicted time instants is shown within the corresponding vorticity profiles.
			The remaining system parameters are the same as in Fig. \ref{fig:6}} \label{fig:8}
\end{figure}    	

\subsubsection{Dynamics of the kinetic energy and the integrated vorticity}\label{vorticity_and_kinetic_conf2}

To further understand the dynamics we next investigate the behavior of the kinetic energy of the system as time evolves, see Fig. \ref{fig:7} $(a)$. 
Note that $E^{\rm kin, ic}_{A}$ is negligible throughout the time evolution, since there is no inherent or emergent vorticity present in species A. 
Indeed, the dynamics of species A is dominated by sound waves as it can be deduced by the finite value of $E^{\rm kin, c}_{A}$. 
However, as discussed in section \ref{vorticity_and_kinetic_conf1}, $E^{\rm kin, ic}_{B}$ increases rapidly after the collision of the two segments signaling the generation of clusters of vortices. 
Also, one can indeed notice the annihilation of a large number of vortices and the production of a stable vortex structure from the time evolution of $E^{\rm kin, ic}_{B}$ 
and $E^{\rm kin, c}_{B}$. 
In particular, the decreasing behavior of $E^{\rm kin, ic}_{B}$ reveals that a large number of vortices annihilates 
until $t = 50$ ms, contributing to an enhancement of the acoustic energy in the system. 
After the final, single vortex dipole remains, the $E^{\rm kin, ic}_B$ oscillates around a mean value with a periodically varying amplitude  
\footnote{It is worthwhile noticing that in the case that the dipole is not rotating within the condensate, then $E^{\rm kin, ic}_B$ should perform 
	an oscillatory motion in time characterized by a constant amplitude. 
	Also we remark that in the absence of an external trap the dipole can not rotate within the condensate.}.
This suggests that the motion of the dipole is affected by both the mutual interaction between the vortices and the spatial inhomogeneity of the condensate. 
Indeed, the vortex and the antivortex propel each other in their direction of flow \cite{Roberts1982}. 
It is also known \cite{Middelkamp2011,Middelkamp2010} that for a VAV in a trap, the spatial inhomogeneity of the density drives each element of the pair to a direction 
which is opposite to their mutually driven motion. 
Depending on the competition between mutually driven motion and the spatial inhomogeneity, diverse trajectories  might result, which essentially 
exhibit a quasi periodic behavior \cite{Middelkamp2011}. 
Hence, the rapid oscillations of $E^{\rm kin, ic}_B$ in time stem from the aforementioned complicated quasi periodic motion of the vortex dipole \cite{Li2008}.  
Figure \ref{fig:8} presents the density and vorticity profiles illustrating the dynamics of the vortex dipole. 
It becomes clear from Figs. \ref{fig:8} ($d_1$)-($d_3$) and ($f_1$)-($f_3$) that the dipole rotates around the trap center, and the distance between the vortex 
and antivortex fluctuates in time indicating their mutual interaction. 
The frequency of the rapid oscillation is determined by the mutual interaction between the pair, while the resulting envelope of the oscillation 
can be attributed to the precession of the pair around the trap center.
Closely inspecting Fig. \ref{fig:7}, it can be observed that whenever the amplitude of oscillation of $E^{\rm kin, ic}_B$ is small (large), a ``chunk'' 
of acoustic energy is emitted in species B (species A), see for example $E^{\rm kin, ic}_B$ at $110$ ms $< t < 165$ ms and $165$ ms $< t < 210$ ms and in particular 
the arrows in Fig. \ref{fig:7}.  
A more detailed study on the dynamics of the vortex dipole can be found in Refs. \cite{Middelkamp2011,Middelkamp2010}. 

The evolution of the integrated vorticity  is also presented in the left inset of Fig. \ref{fig:7}. 
At the early stages of the dynamics, $0$ ms $< t < 10$ ms, the integrated vorticity exhibits the same behavior as in section \ref{vorticity_and_kinetic_conf1}.  
Namely, both $\mathcal{Z}_A$ and $\mathcal{Z}_B$ increase after the quench, reaching a maximum value and then show a decreasing behavior in time. 
However there are few noticeable differences. 
For instance, the maximum value of both the $\mathcal{Z}_A$ and the $\mathcal{Z}_B$ is smaller than their counterparts corresponding to the configuration 
with an initially singly quantized vortex imprinted in species A. 
Moreover, during the entire time evolution $\mathcal{Z}_A$ and $\mathcal{Z}_B$ remain quite disparate in their values. 
Indeed the overall vorticity of species A and species B is different since a vortex dipole is present in species B while 
no vortex is generated in species A. 
The behavior of the integrated vorticity of species A is dominated only by the acoustic waves, whereas that of species B is associated with both the 
vorticity and the acoustic waves.  
Finally, the angular momentum of both species remains close to its initial value throughout the evolution, see the right inset in Fig. \ref{fig:7}. 

For reasons of completeness let us also briefly comment on the effect when an initially zero phase 
difference between the segments of species B occurs and species A possesses zero vorticity (results not show here for brevity). 
Quenching the amplitude of the potential barrier of species B induces a counterflow dynamics of its segments and as a consequence dark and bright stripes 
form in species B and species A respectively. 
Then, the dark stripes formed in species B bend and two pairs of vortex dipoles are nucleated which evolve in time and subsequently 
one antivortex annihilates at the edge of the condensate resulting in the formation of a VAV-AV structure. 
As time evolves one of the participating antivortices of the VAV-AV structure moves out of the condensate 
resulting in a vortex dipole which undergoes a similar quasi periodic motion as discussed earlier in Sec. \ref{Discussion2} and eventually disappears. 
During the above-described dynamics of species B, the bright solitons formed in species A always accompany the vortex structures present in species B and of course 
reside at the same place. 
Moreover, the angular momentum of each of the species remains close to zero during the entire time evolution and therefore no transfer of angular momentum occurs.

\section{Dynamics of a strongly mass imbalanced mixture}\label{unequal_Mass}

Next we unravel the quench induced nonequilibrium dynamics of a binary condensate consisting of strongly imbalanced atomic species. 
Again, our focus is the initial configuration where species B contains a relative phase difference of $\pi$ between its segments and species A 
possesses a singly quantized vortex. 
Exemplarily, we consider the experimentally realizable mixture consisting of $^{87}$Rb-$^{133}$Cs species where $m_r \approx 0.65$. 
In the following, for convenience, we label $^{87}$Rb as species A and $^{133}$Cs as species B.  
Moreover, the intraspecies scattering length of the $^{87}$Rb and $^{133}$Cs atoms correspond to $100.4a_0$ and $280a_{0}$~\cite{McCarron2011} respectively, 
whereas the interspecies scattering length is fixed to the value $100a_{0}$. 
Furthermore, we use a harmonic oscillator potential of frequency, $\omega_A = \omega_B = 2\pi \times 30.832Hz$ while the 
typical anisotropy parameters are taken to be $\alpha = 1$ and $\lambda = 40$ \cite{Vogels2001}. 
As in the previous section, after preparing the system in its initial state we suddenly ramp down the potential barrier in species B at $t=0$ 
and monitor the dynamics of the system.
\begin{figure}[ht]
	\centering
	\includegraphics[width=0.42\textwidth]{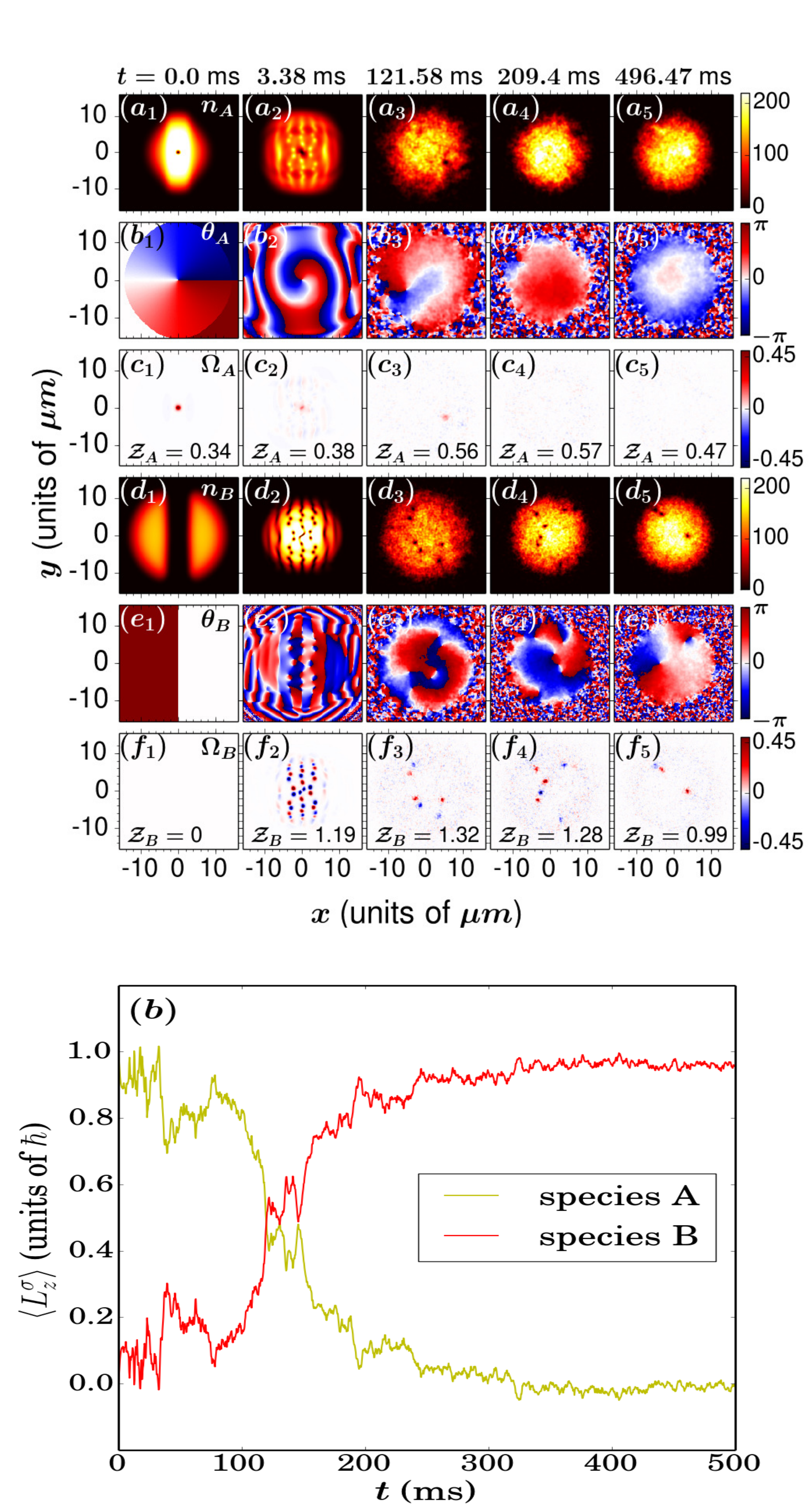}
\caption{(Color online) Density profiles of ($a_1$)-($a_5$) species A ($^{87}$Rb) and ($d_1$)-($d_5$) species B ($^{133}$Cs) at selected 
time instants of the evolution. Also shown are the corresponding vorticity [phase] profiles of ($c_1$)-($c_5$) [($b_1$)-($b_5$)] species A and ($f_1$)-($f_5$)[($e_1$)-($e_5$)] 
species B. The  integrated vorticity of each species at these time instants are shown in the corresponding profiles.
The binary BEC consists of $N_A=N_B=10^4$ bosons and it is prepared in its ground state with a singly quantized vortex in species A and a phase 
difference in species B such that $\theta_B(x) = \pi$ for $x < 0$ and $\theta_B (x) = 0$ for $x \geq 0$. 
The intraspecies scattering lengths are $a_{AA} = 100.4a_0$ and $a_{BB} = 280.0a_0$, while the interspecies scattering length corresponds to $a_{AB} = 100.0a_0$. 
(b) Dynamics of the angular momentum $L^{\sigma}_z(t)$ of the $\sigma = A, B$ species along the $z$ direction.}
\label{fig:CS}
\end{figure} 

The emergent density and vorticity profiles of both species at different time-instants, following a quench of the potential barrier 
of species $B$, are presented in Fig. \ref{fig:CS}. 
A close inspection of the initial state [Figs. \ref{fig:CS} ($a_1$), ($c_1$)] reveals that for the $^{87}$Rb-$^{133}$Cs mixture the interface between the 
two species is flatter than that of the equal mass scenario [compare Fig. \ref{fig:CS} ($c_1$) and Fig. \ref{fig:1} ($c_1$)]. 
This is due to the fact that while the presence of the potential barrier at the trap center prevents species B to occupy this spatial region, simultaneously 
the system desires to minimize its total energy by placing the heavier atoms close to the trap center. 
Directly after the barrier ramp down, the overall dynamics taking place is similar to the equal mass case, namely we observe the formation of V-AV pairs 
[Figs. \ref{fig:CS} ($a_3$), ($b_3$), ($c_3$), ($d_3$)] 
via the snake instability, see Figs. \ref{fig:CS} ($a_2$), ($c_2$) and Figs. \ref{fig:CS} ($b_2$), ($d_2$) for the underlying phase profiles.  
Notice also that at the early stages of the dynamics [Figs. \ref{fig:CS} ($c_2$)] a larger number of vortices is generated in species B which is composed of heavier atoms 
when compared to the almost mass balanced case [Fig \ref{fig:1}($c_5$)].  
Although these non-linear excitations interact and gradually decay they appear to be more robust [Fig. \ref{fig:CS}($c_4$)] compared to the equal mass scenario [Fig. \ref{fig:2}($c_7$)] 
in the long-time dynamics of the system. 
This behavior is attributed to the fact that $^{133}$Cs has a smaller healing length than $^{85}$Rb since the former species is heavier and has a larger scattering length 
than the latter. 
Moreover, the vortices building upon species B and the corresponding bright solitons formed in species A during the dynamics exhibit a random motion 
in a very asymmetrical manner with respect to the trap center, see Figs. \ref{fig:CS}($c_4$) and ($c_5$). 
Phrased otherwise, the density humps (bright solitons) in species A are distributed non-uniformly with respect to the position of the initially imprinted vortex  
which in turn experiences a net velocity field created by the bright solitons. 
As a result, the initially imprinted vortex in species A is displaced from the trap center moving towards the edge of the condensate while performing a precessional 
motion [Fig. \ref{fig:CS}($c_3$)] and finally drifts out of the condensate [see Figs. \ref{fig:CS}($a_3$) and ($a_4$)]. 
Indeed, as shown in Fig. \ref{fig:CS}($c_4$) at $t = 496.47$ ms there is a single V-AV pair located at the edge of the condensate and one vortex residing 
relatively close to the trap center. 
Concluding, the resultant spontaneously generated nonlinear structures appear to be more robust when species B possess a significantly larger mass 
than species A. 
Moreover, during the above-mentioned process, an ultimate transfer of angular momentum from species A to species B occurs, see Fig. \ref{fig:CS}($b$). 
Therefore we can deduce that even for the strongly mass imbalanced case the generic phenomenon of angular momentum transfer between the species takes place. 
Most importantly, for a strongly imbalanced mixture the angular momentum transfer is favored to a larger extent since the heavier species exhibits a smaller healing length 
and as a consequence it is possible to accommodate a larger amount of vortices. 
Indeed, as can be seen in Fig. \ref{fig:CS}($b$) the angular momentum of species B gradually reaches unity. 

Finally, let us comment that for a lighter species B, for instance, assuming a $^{87}$Rb-$^{23}$Na binary BEC~\cite{Wang2015}, 
due to the larger healing length of the $^{23}$Na component, not only the number of the generated V-AV pairs decreases significantly, but they  also decay 
rapidly during the dynamics (results not shown for brevity).

\section{Conclusions}\label{conclusion}

We have investigated the spontaneous generation of vortex-bright-solitons induced by merging two initially separated segments of one of the species in a two-dimensional 
binary BEC, considering both slight and considerable mass imbalance between the two species. 
The segmented species possess initially a $\pi$ phase difference between its segments while in the other species a singly quantized vortex or no vortex
is imprinted. 
The merging of the segmented species is triggered by removing the potential barrier that initially separates the two segments of one of the species. 
We have described the outcome of the interference of the two initially separated segments and their effects on the other species. 
In particular, we have monitored the generation of vortex complexes, their motion and interactions within the clouds, and the resulting vortex-bright-soliton structure 
formation at long-time scales. 
The impact of the interspecies repulsion on the quench induced vortex generation and the dynamics of the angular momentum of both species have also been examined. 
To gain insights into the dynamics of the system, we have utilized several diagnostics such as the vorticity, the $z$ component of the angular momentum, the integrated vorticity 
and the compressible and incompressible parts of the kinetic energy. 

In the first part of our study we considered a binary BEC with slightly mass imbalanced components. 
The initial configuration consists of a segmented species with a $\pi$ relative phase difference 
between the segments and a singly quantized vortex imprinted on the other species. 
A major finding here was the ultimate relaxation of
species B into a state with a singly quantized vortex, caused by the angular momentum transfer from 
species A to species B. An accompanying bright soliton results
in species A with the emerging vortex-bright (or,more
  precisely vortex-antidark)
structure persisting for long times. 
In particular, after merging the two segments of species B we observe the generation of dark and bright soliton stripes in species B and species A respectively. 
These stripes break into VAV pairs \cite{Feder2000} via the manifestation of the snake instability and subsequently form a VAV-V structure \cite{Seman2010} which evolves 
in the long-time dynamics into vortex-bright solitons. 
The overall dynamics can also be understood by inspecting the incompressible and compressible parts of the kinetic energy of the system 
being associated with the existence of vortices and sound waves respectively. 
Moreover, the transfer rate of the angular momentum between the species is found to crucially depend on the value of the interspecies interactions. 
More specifically for larger interspecies interactions, within the miscible regime of interactions, an increase of the transfer rate occurs while in the phase separated regime 
both species tend to share the same angular momentum. 

As a next step, we consider an initial configuration in which a $\pi$ relative phase difference occurs between the segments of species B, 
but species A contains zero vorticity. 
As in the above-mentioned scenario the collision of the two segments of species B gives rise to dark and bright soliton stripes in both species which 
are subjected to the consequent modulational instability and break into vortex-antivortex pairs. 
In contrast to the previous case these vortex-antivortex pairs evolve into quadruples \cite{yang2016dynamics} resulting in the long-time dynamics into 
a vortex dipole in species B accompanied by two bright solitons in species A. 
Another distinct feature of this configuration is the absence of angular momentum transfer between the species. 
Next, we inspected the effect of a heavier segmented species possessing also an experimentally relevant larger intraspecies scattering length. 
As a result the healing length of this heavier species becomes reduced compared to the mass balanced scenario and favors a complete transfer of angular momentum 
from species B to species A within the miscible regime of interactions.  
Moreover, the same overall phenomenology to the mass balanced case takes place during the dynamics while the quench-generated non-linear structures become more robust and 
persist in the long-time dynamics of the system. 
In other words, our computations suggest that the mass imbalance mainly affects the time scale of the appearance of the relevant phenomenology.

Summarizing, we can conclude that the robustness of the quench-induced topological structures and as a consequence the overall vortical activity of the binary 
condensate depends strongly on the mass imbalance between the two species. 
In particular, a slightly mass imbalanced mixture with a $\pi$ phase difference between the segments of species B and a singly quantized vortex 
imprinted in species A, leads ultimately to the persistence of a single vortex of unit charge in species B. 
However, with similar initial phases, increasing the mass of the species B to a fairly large value enhances the robustness of the quench-induced generated structures. 
In both cases, it should be noted that the structures are coupled to
a bright (or more accurately anti-dark as it sits on top of a
background~\cite{danaila,kiehn2019spontaneous}) soliton in the other species.

There are several interesting research directions to consider in future endeavors. 
A straightforward one is to examine the dynamical formation of vortex complexes when considering multiple
fragments which possess an arbitrary relative phase difference 
between each other and unravel the corresponding quench dynamics upon ramping down the barriers that initially separate the fragments. 
Another interesting prospect would be to examine the dynamics of a harmonically trapped binary BEC where one of its components contains initially a multicharged vortex. 
Here, we have considered the case where the second species has a relative phase
of $\pi$, however a more systematic study as a function of the relative
phase of the two fragments in that species would be of interest in its
own right. Moreover, the studies of the dynamics and interactions of these emerging vortex-bright (or vortex antidark)
patterns that arise in these numerical experiments are equally
interesting. Furthermore, this is the case
both for the setting of same, as well as for that of opposite charge
vortices. 
These investigations could also be extended to spin-orbit coupled BECs \cite{Stanescu2008, Wang2010, spielman} and dipolar BECs \cite{Santos2002,Lahaye_2009}, where one can further 
inspect the impact of the additional interaction term among the particles on the vortex generation. 
The above-mentioned studies would also be equally interesting at the beyond mean-field \cite{Katsimiga2017, Katsimiga_2017, Katsimiga2018,katsimiga2018many} level where intra- and interspecies 
correlations are taken into account.

\begin{acknowledgments}
K.M. acknowledges a research fellowship (Funding ID no 57381333)  from  the Deutscher Akademischer Austauschdienst (DAAD).  K.M. thanks S. Majumder, 
D. Angom and S. Bandopadhyay for helpful discussions. 
This material is also based upon work supported by the U.S. National Science Foundation under Grant No. PHY-1602994 (P.G.K.).
\end{acknowledgments}

\appendix

\section{Dynamics of initially non-symmetric segments in the double-well}\label{Imperfection}

\begin{figure}[ht]
	\centering
	\includegraphics[width=0.43\textwidth]{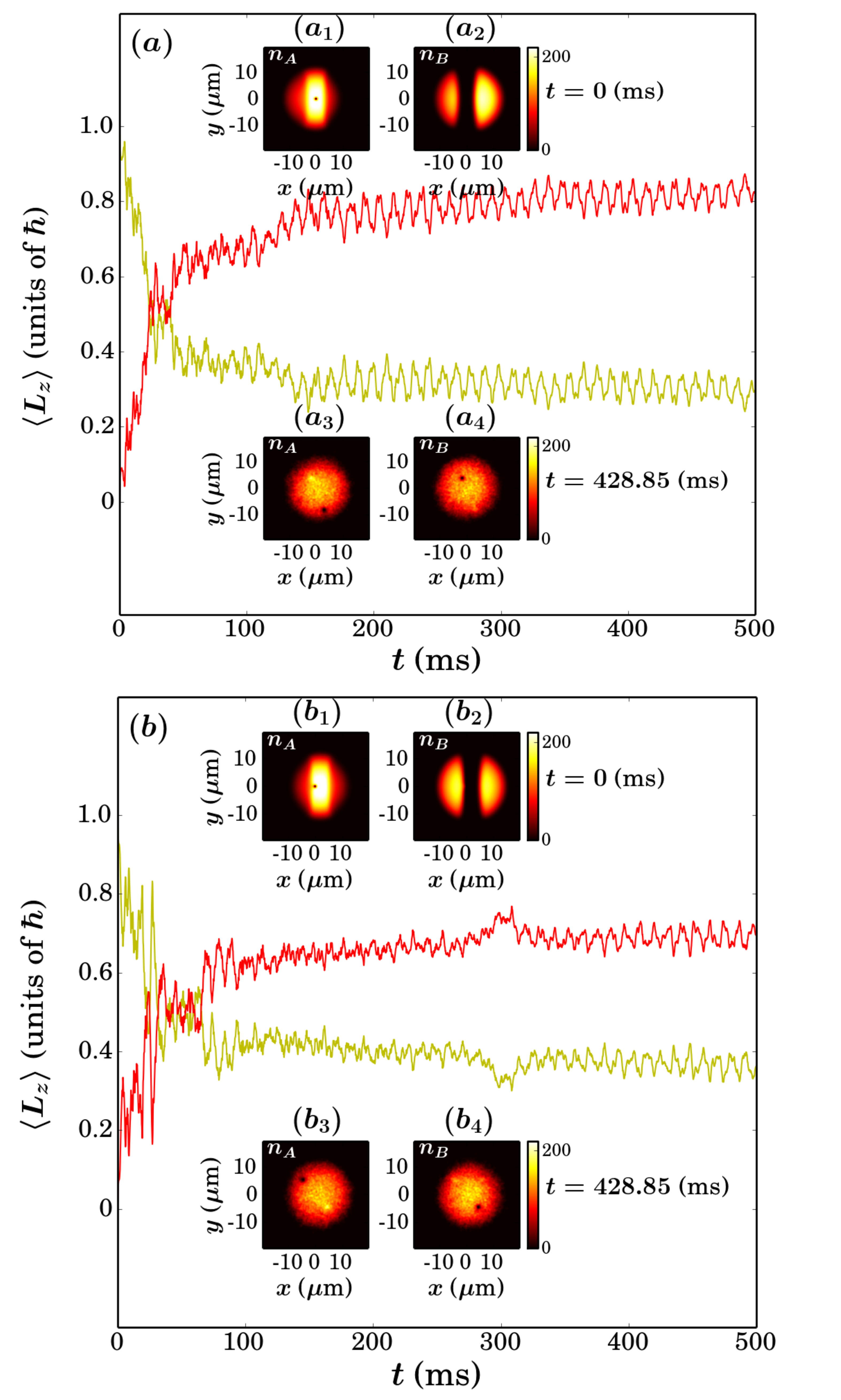}
	\caption{($a$) Temporal evolution of the $\sigma = A, B$ species angular momentum $L^{\sigma}_z (t)$ along the $z$ direction when there 
	is an initial particle imbalance between two segments. Snapshots of the $\sigma$ species density (see legends) at ($a_1$)-($a_2$) the initial time ($t = 0$ ms) 
	and ($a_3$)-($a_4$) the long-time ($t = 428.85$ ms) dynamics of the system. ($b$) Dynamics of the angular momentum $L^{\sigma}_z(t)$ along the $z$ direction when the two 
	segments of species B are not symmetrically located with respect to the trap center. Density snapshots of the $\sigma$ species at ($b_1$)-($b_2$) the initial 
	time ($t = 0$ ms) and ($b_3$)-($b_4$) the long-time dynamics ($t = 428.85$ ms). } 
	\label{fig:imp}
\end{figure}

Here, we discuss the effect of an initial particle imbalance between the two segments of species B on the quench 
induced structure formation. 
As previously [Sec. \ref{Discussion1}], species A is confined in an isotropic 2D harmonic oscillator and species B in an anisotropic trap composed of a harmonic 
oscillator of equal frequencies in both directions and a potential barrier in the $y$ direction. 
The intra- and interspecies scattering lengths correspond to $a_{AA} = 100.4a_0$, $a_{BB} = 95.0a_0$ and $a_{AB} = 60a_0$ respectively. 
Also, the mixture is prepared such that the B species has a $\pi$ phase difference between its segments and species A contains a singly quantized vortex [Fig.~\ref{fig:imp}($a_1$)]. 
However, in the present case we additionally impose a linear tilt of the form $-dx+c$ with $d=1$ and $c=10$ in the trap of the B species. 
This linear tilt essentially causes an energy offset between the left ($x<0$) and right ($x>0$) spatial regions of the  B species potential resulting in 
a particle imbalance between the segments of species B. 
Consequently due to the particular choice of the parameters $d$, $c$ the left segment of species B possesses $25\% N_B$ while the right segment 
has $75\% N_B $ [Fig.~\ref{fig:imp} ($a_2$)]. 
As a result of the particle imbalance between the segments of species B and the finite positive value of $a_{AB}=60a_0$ also the density of species A exhibits initially 
an asymmetric configuration with respect to $x=0$ [Fig.~\ref{fig:imp} ($a_1$)]. 
In particular less particles of species A are located at $x>0$ since the atoms of species B are predominantly residing in this spatial direction ($x>0$) due to the presence 
of the tilt [Fig.~\ref{fig:imp} ($a_1$)] and subsequently repel the particles of species A to $x<0$. 
To induce the dynamics we then ramp down the potential barrier of species B and let the system to evolve. 
Monitoring the time-evolution of the $\sigma$ species angular momentum we observe that it is gradually transferred from species A 
to species B, see Fig.~\ref{fig:imp} ($a$). 
Closely inspecting $L^{\sigma}_z(t)$ we deduce that its transfer rate is maximized within the interval $0<t<100$ ms. 
Most importantly, the angular momentun transfer between the species occurs faster for this particle-imbalanced case of species B as compared to 
the particle balanced scenario discussed in Sec. \ref{Discussion1} and presented in Fig. \ref{fig:4} ($c$). 
Another noticeable difference in the behavior of $L^{\sigma}_z(t)$ between the particle balanced [Fig.~\ref{fig:4}($c$)] and imbalanced [Fig.~\ref{fig:imp} ($c$)] case 
is that $L^{\sigma}_z$ exhibits a more pronounced oscillatory behavior in the latter case and also the mean value of $L^{A}_z(t)$ [$L^{B}_z(t)$] 
in the long-time dynamics is larger [smaller]. 
However, as in the particle balanced case, each of the species contains a single vortex in the long-time dynamics of the system, see e.g. Figs.~\ref{fig:imp}($a_3$)-($a_4$). 
This clearly reveals that the non-linear structure formation in the long-time dynamics depends mainly on the phase difference between the segments of species B, while a particle imbalance 
between these segments only alters the time scale of the manifestation of the relevant phenomenology. 

Next, we comment on the impact of another type of initial state imperfection on the quench dynamics of the mixture. 
Here, the initial state preparation is the same as in Sec. \ref{Discussion1} but the two segments of species B are not symmetrically placed with respect to the trap 
center $x = 0$, see Fig.~\ref{fig:imp} ($b_2$). 
This can be achieved by considering a displaced potential barrier namely $V_L = \beta(t)U_0e^{-\frac{(x -x_0)^2}{2a^2}}$, where $x_0 = 1 a_{\rm osc}=1.94 \mu m$, 
$\beta(t=0)=1$, $U_0=20\omega_x$ and $a=1a_{osc}$. 
As it can be seen, the vortex in species A is located at the trap center [Fig.~\ref{fig:imp}($b_1$)] while the overall density of species A is somewhat 
shifted to $x>0$ due to the location of the potential barrier in species B at $x_0$. 
The dynamics of the $\sigma$ species angular momentum is illustrated in Fig. \ref{fig:imp} ($b$) following a quench of the potential barrier height to zero. 
Evidently, an angular momentum transfer between the species occurs with a maximum transfer rate taking place within the interval $0<t<100$ ms. 
Interestingly, the final value of $L^{\sigma}_z$ is very close to the one corresponding to the symmetrically located 
segments, compare Fig.~\ref{fig:imp}($b$) and Fig.~\ref{fig:4}($c$). 
Moreover, we can infer that this initial state imperfection again modifies the time scales of the dynamical response of the mixture. 
This alteration of time scales can be directly observed in the dynamics of $L^{\sigma}_z$ between the case of $x_0=0$ [Fig. \ref{fig:4} ($c$)] 
and $x_0=1 a_{\rm osc}$ [Fig. \ref{fig:imp} ($b$)] 
However, in the long-time dynamics of the system only a single vortex persists in both species, see Figs. \ref{fig:imp} ($b_3$), ($b_4$). 
Concluding, also this initial state imperfection does not alter both the non-linear structure formation in the long-time dynamics and the underlying transfer of 
angular momentum between the species.

\section{Derivation of the equations of evolution of the angular momentum}\label{Derivation}

In this Appendix we provide the derivation of the equations of
temporal evolution of the angular momentum 
of each species [see Eqs. (\ref{9}) and (\ref{10})] used in the main text to explain the angular 
momentum transfer between the species during the nonequilibrium dynamics of the binary BEC. 
By invoking its definition [see Eq. (\ref{5})], the $z$ component of the angular momentum of 
the $\sigma$ species reads  
\begin{equation}
L^{\sigma}_z = -i \int \dd S \psi^{*}_{\sigma}(x \partial_y - y\partial_x)\psi_{\sigma},\label{ang_mom}
\end{equation}
where $\dd S = \dd x \dd y$ denotes the spatial area of integration. 
In our case it corresponds to the $x$-$y$ plane restricted by the hard-wall boundaries used herein, i.e. $x_{\pm}=y_{\pm}=\pm10$. 
To proceed, we introduce the abbreviation $\tilde{L} = x \partial_y - y \partial_x$. 
Then we can express the angular momentum operator as $L_z = -i\tilde{L}$. 
Since $L_z$ is a self-adjoint operator we obtain also that $\tilde{L}^{\dagger} = -\tilde{L}$. 
Calculating the time-derivative of Eq. (\ref{ang_mom}) and also using the above-mentioned abbreviation we arrive at the relation 
\begin{equation}
\begin{split}
\frac{\dd L^{\sigma}_z}{dt} = &-i\int \dd S\frac{\partial \psi^{*}_{\sigma}}{\partial t}\tilde{L}\psi_{\sigma}  -i \int \dd S \psi_{\sigma} \tilde{L} \frac{\partial \psi^{*}_{\sigma}}{\partial t} 
\\&= -i\int \dd S \frac{\partial \psi^{*}_{\sigma}}{\partial t}\tilde{L}\psi_{\sigma} + i \int \dd S \frac{\partial \psi}{\partial t} \tilde{L}\psi^{*}_{\sigma}.\label{der}
\end{split}
\end{equation}

Subsequently, we replace the time-derivative of the $\sigma$ species wavefunction using the coupled set of GP equations 
[Eqs. (\ref{3}) and (\ref{4}) in the main text]. 
The resulting time-derivative of the angular momentum [Eq. (\ref{der})] reads  
\begin{equation}
\begin{split}
\frac{d L^{\sigma}_z}{dt} &= \int dS (-\frac{1}{2}\nabla^{2}_\perp\psi^{*}_{\sigma} + V_{\sigma} \psi^{*}_{\sigma} + \mathcal{N}_{\sigma}\psi^{*}_{\sigma})\tilde{L}\psi_{\sigma} 
\\&+ \int dS \tilde{L} \psi^{*}_{\sigma}(-\frac{1}{2}\nabla^{2}_\perp\psi_{\sigma} + V_{\sigma} \psi_{\sigma} + \mathcal{N}_{\sigma}\psi_{\sigma}) 
\end{split}
\end{equation}
Furthermore by substituting $L_z = -i\tilde{L}$ into the last equation and performing some algebra we obtain the following equation of motion 
of the $\sigma$ species angular momentum 
\begin{equation}
\begin{split}
&\frac{d L^{\sigma}_z}{dt} \\& =\frac{1}{2}\int dS \big [ \nabla^{2}_\perp \psi^{*}_{\sigma}(x\frac{\partial}{\partial y} - y\frac{\partial}{\partial x})\psi_{\sigma} + \nabla^{2}_\perp \psi_{\sigma}
(x\frac{\partial}{\partial y} - y\frac{\partial}{\partial x})\psi^{*}_{\sigma} \big ]  \\
& + \int dS \big [ V_{\sigma} \psi^{*}_{\sigma}(x\frac{\partial}{\partial y} - y\frac{\partial}{\partial x})\psi_{\sigma} dS  +  V_{\sigma} \psi^{*}_{\sigma}(x\frac{\partial}{\partial y} - y\frac{\partial}{\partial x})\psi_{\sigma}  ] \\
& + \int dS \big [ \mathcal{N}_{\sigma} \psi^{*}_{\sigma}(x\frac{\partial}{\partial y} - y\frac{\partial}{\partial x})\psi_{\sigma} dS  +  \mathcal{N}_{\sigma} \psi^{*}_{\sigma}(x\frac{\partial}{\partial y} - y\frac{\partial}{\partial x})\psi_i  ]. \label{time_der_ang}
\end{split}
\end{equation}\\
Here, $\mathcal{N}_{\sigma} = \sum_{\sigma' }^{}\mathcal{U}_{\sigma \sigma'}\abs{\psi_{\sigma}}^2$ denotes the nonlinear interaction term.
As it can readily seen by inspecting Eq. (\ref{time_der_ang}) there three different contributing terms, corresponding to the three different lines in Eq. (\ref{time_der_ang}), 
in the equation of motion of the angular momentum. 
It can be shown through a tedious but straightforward calculation that the first term associated with the kinetic energy of the system vanishes.  
The second and third terms related to the potential energy and the nonlinear interactions respectively reduce to expressions 
$-\int \dd S \abs{\psi_{\sigma}}^2(x\frac{\partial}{\partial y} - y\frac{\partial}{\partial x})V_{\sigma}$ and 
$-\int \dd S \abs{\psi_{\sigma}}^2(x\frac{\partial}{\partial y} - y\frac{\partial}{\partial x})\mathcal{N}_{\sigma}$. 
As a consequence the equation of motion of the angular momentum of the $\sigma$ species reads
\begin{equation}
\begin{split}
\frac{d L^{\sigma}_z}{dt} = &-\int dS \abs{\psi_{\sigma}}^2(x\frac{\partial}{\partial y} - y\frac{\partial}{\partial x})V_{\sigma} \\& -\int dS \abs{\psi_{\sigma}}^2(x\frac{\partial}{\partial y} 
- y\frac{\partial}{\partial x})\mathcal{N}_{\sigma}.
\end{split}
\end{equation}
These correspond exactly to the Eqs. (\ref{9}) and (\ref{10}) presented in the main text.


\bibliographystyle{apsrev4-1}
\bibliography{bibliography}

\end{document}